\documentclass[
               reprint, twocolumn,
               preprintnumbers,
               amsmath,
               amssymb,
               showpacs,
               aip,jcp,
               superscriptaddress]{revtex4-1}

 \usepackage{ucs}
 \usepackage[utf8x]{inputenc}
 \usepackage{latexsym}
 \usepackage{amsfonts}
 \usepackage{amsmath}
 \usepackage[dvips]{graphicx}
 \usepackage{color}
 \usepackage{bm}
 \usepackage{hyperref}
 \usepackage[ruled, lined, longend]{algorithm2e}
 \usepackage{xcolor}
\hypersetup{
   colorlinks=true,       
   linkcolor=magenta,          
   citecolor=magenta,        
   filecolor=magenta,      
   urlcolor=blue           
}

\begin{document}

\author{Sudip Sasmal}
\email[e-mail: ]{sudip.sasmal@pci.uni-heidelberg.de}
\affiliation{Theoretische Chemie,
             Physikalisch-Chemisches Institut,
             Universität Heidelberg,
             Im Neuneheimer Feld 229, 69120 Heidelberg, Germany}

\author{Oriol Vendrell}
\email[e-mail: ]{oriol.vendrell@pci.uni-heidelberg.de}
\affiliation{Theoretische Chemie,
             Physikalisch-Chemisches Institut,
             Universität Heidelberg,
             Im Neuneheimer Feld 229, 69120 Heidelberg, Germany}
\affiliation{Interdisciplinary Center for Scientific Computing,
             Universität Heidelberg,
             Im Neuneheimer Feld 205, 69120 Heidelberg, Germany}

\title{Sum-of-products form of the molecular electronic Hamiltonian and
       application within the MCTDH method}

\date{\today}

\begin{abstract}
We introduce two different approaches to represent the second-quantized electronic Hamiltonian in
a sum-of-products form. These procedures aim
at mitigating the quartic scaling of
the number of terms in the Hamiltonian
with respect to the number of spin orbitals,
and thus enable applications to larger molecular systems.
Here we describe the application of these approaches within the
multi-configuration time-dependent Hartree framework.
This approach is applied to the calculation of eigen energies of LiH and 
electronic ionization spectrum of H$_2$O.
\end{abstract}

\maketitle

\section{Introduction} \label{sec:introduction}
\par
The multiconfiguration time-dependent Hartree (MCTDH)
~\cite{mey90:73,man92:3199,bec00:1,mey03:251,mey09:book}
and its multilayer generalization
(ML-MCTDH)~\cite{wan03:1289,man08:164116,Ven11:44135,mey12:351,wan15:7951}
are very efficient methods to simulate high dimensional quantum dynamics of
nuclear degrees of freedom. In its original form,
the MCTDH \emph{Ansatz} cannot describe a system of indistinguishable
particles as it assumes a product form of the underlying SPFs and thus,
does not reflect the proper symmetry of the indistinguishable particles.
However, one can construct the multiconfiguration wavefunction in the basis of
Slater determinants and permanents to treat systems of fermions and
bosons, respectively. These theories are called MCTDH for fermions
(MCTDH-F)~\cite{cai05:12712,alo07:154103,hoc11:084106,sat13:023402,lod20:011001}
and bosons
(MCTDH-B)~\cite{alo07:154103,alo08:33613}.
An unified
version of the two theories using a non-symmetric core tensor to connect
mixtures of different types of
indistinguishable particles has also been established~\cite{kroe13:63018}.
A limitation of such descriptions is that the number of configurations
of the electronic/bosonic subsystem increases combinatorially with the
number of particles and single-particle functions, while 
the antisymmetry/symmetry requirement prevents their further decomposition
into smaller-rank tensors.
Hence, within the same type of indistinguishable particle, MCTDH-B and MCTDH-F approaches
are incompatible with the multilayer extension of the MCTDH framework.

A fundamentally different alternative to describe systems of indistinguishable
particles is to use the second quantization representation. Wang and Thoss
described and applied this approach in the context of MCTDH and called it
MCTDH in SQR (MCTDH-SQR)~\cite{wan09:024114}.
Here, the state of the system is described in
the occupation number representation referring to the occupation
of a given set of spin-orbitals. The symmetry of the
indistinguishable particles is encoded in the creation and annihilation operators
acting on the state of the system.
In this approach, as the coordinates are the occupation of individual spin-orbitals,
the DOFs are distinguishable and hence a multilayer \emph{Ansatz} of the wavefunction
is straightforward.
This approach has been used in a number of applications after its introduction, e.g.
to solve the impurity problem that appears in non-equilibrium
dynamical mean-field theory~\cite{Bal15:45136},
quantum transport in molecular junctions~\cite{Wan11:244506,wan13:134704,wan13:7431}
and quantum dots~\cite{Wil13:045137,Wil14:205129}.
Manthe and Weike developed an MCTDH-SQR approach based on time-dependent
optimal (spin-)orbitals, the MCTDH-oSQR method~\cite{man17:064117,wei20:034101}.
The aforementioned applications were concerned with model
Hamiltonians~\cite{wan09:024114,Bal15:45136,man17:064117,wei20:034101,Wan11:244506,wan13:134704,wan13:7431,Wil13:045137,Wil14:205129}.
Recently, we extended the MCTDH-SQR method to describe the non-adiabatic dynamics
in molecular systems based on the second-quantized representation of the electrons
and first-quantized representation of the nuclear coordinates while providing
expressions for the non-adiabatic coupling matrix elements in this combined
representation~\cite{Sas20:154110}.
In this formalism, most non-adiabatic effects are described by the time-evolution
of the electronic subsystem coupled to the dynamics of the nuclei and bypasses
the explicit calculation of non-adiabatic couplings in terms of electronic states.
Thus, the approach provides an alternative to the usual group BO approximation.

However, the major problem of the MCTDH-SQR method applied to \emph{ab initio}
studies of large molecular
systems remains the enormous size of the electronic SQR Hamiltonian, whose number of
terms increases with the fourth power of the number
of spin-orbitals. The density matrix
renormalization group (DMRG)~\cite{Whi92:2863,Whi93:10345,sch05:259}
formalism, which is based on a similar tensor representation compared
to ML-MCTDH, avoids such scaling by representing
the Hamiltonian as matrix product operator
(MPO)~\cite{Sch11:96,Cha11:465,McC07:P10014,Cha16:014102,Kel15:244118,Yan14:283}
that matches the matrix
product state (MPS)~\cite{Sch11:96,McC07:P10014} structure of the wavefunction.
A comparably compact form of the electronic Hamiltonian based on a sum-of-products (SOP) expansion, which is able to regain the maximum performance from the multiconfigurational form of the MCTDH
wavefunction, has not yet been put forward.
In comparison, for the nuclear dynamics problem, there exist a plethora of methods to bring general potential energy surfaces to a SOP form, including the
\texttt{POTFIT}~\cite{bec00:1,jaec96:7974,jaec98:3772},
multigrid \texttt{POTFIT}~\cite{Pel13:014108},
Monte Carlo \texttt{POTFIT}~\cite{Sch17:064105}
and its multilayer variant~\cite{Ott18:116},
multilayer \texttt{POTFIT}~\cite{ott14:14106},
Monte Carlo \texttt{CANDECOMP}~\cite{Sch20:024108}
algorithms
and neural network approaches~\cite{Man05:5295,Man06:084109,Man06:194105,Koc14:021101,
She15:144701,Pra16:174305,Pra16:158}.

In this work, we introduce and benchmark two SOP-based strategies that aim at mitigating
the quartic scaling of the electronic Hamiltonian with respect to
the number of spin-orbitals.
The paper is organized as follows.
Section~\ref{sec:theory:sop} reviews the general strategy to write the SQR electronic
Hamiltonian in SOP form.
Section~\ref{sec:theory:dof} discusses the choice of DOF in the MCTDH-SQR
formalism and
Section~\ref{sec:theory:compact} introduces two strategies that lead to
compact SOP forms of the electronic SQR Hamiltonian.
Secs~\ref{sec:res:lih} and ~\ref{sec:res:h2o}
present and discuss numerical results of the LiH and H$_2$O
systems, respectively.
Finally a summary and conclusions are provided in Section~\ref{sec:conclusions}.

\section{Theory}            \label{sec:theory}
\subsection{Sum-of-products form of the electronic Hamiltonian}  \label{sec:theory:sop}
The molecular electronic Hamiltonian in the second quantization framework reads
\begin{align}
 \hat{H} = \sum_{ij} h_{ij} \hat{a}_i^{\dagger}\hat{a}_j + \frac{1}{2}
    \sum_{ijkl}
    v_{ijkl}\hat{a}_i^{\dagger}\hat{a}_j^{\dagger}\hat{a}_l\hat{a}_k ,
    \label{eq:ham_spinorb}
\end{align}
where
\begin{align}
  h_{ij} = \langle \phi_i(1)| -\frac{1}{2}\nabla_1^2 - \sum_{A=1}^M \frac{Z_A}{r_{_{1A}}} | \phi_j(1) \rangle , \\
  v_{ijkl} = \langle \phi_i(1) \phi_j(2) | \frac{1}{r_{12}} | \phi_k(1) \phi_l(2) \rangle ,
\end{align}
are the one- and two-body integrals, involving the
spin orbitals $\phi_i$, respectively.
The $\hat{a}_i$ and $\hat{a}_i^{\dagger}$ correspond to the annihilation
and creation operators that annihilates and creates an electron in the $i$-th
spin orbitals, respectively, satisfy the fermionic commutation relations
\begin{align}
 \label{eq:anticom}
 \{\hat{a}_i, \hat{a}_j^{\dagger} \} =
    \hat{a}_i\hat{a}_j^{\dagger} + \hat{a}_j^{\dagger}\hat{a}_i 
    = \delta_{ij} , \\
 \label{eq:antic4}
  \{\hat{a}_i^{\dagger}, \hat{a}_j^{\dagger} \} = \{\hat{a}_i, \hat{a}_j \} = 0.
\end{align}
Although the electronic Hamiltonian written in Eq.~\ref{eq:ham_spinorb} looks like
already being in a SOP form, the anti-commutation relations of the fermionic creation and 
annihilation operators in Eq.~\ref{eq:anticom} lead to
the accumulation of a phase factor $S_s = \sum_{k=1}^{s-1} n_k$ depending
on the occupation of all spin-orbitals before the $s$-th position
for $\hat{a}_s$ acting on a Fock-space configuration
(the same is true for $\hat{a}_s^\dagger$)~\cite{Fetter2003}
\begin{align}
    \label{eq:opfermi}
    \hat{a}_s |n_1, n_2, \dots , n_M \rangle  =
            \hat{a}_s (\hat{a}_1^\dagger)^{n_1}
                      (\hat{a}_2^\dagger)^{n_2} \cdots
                      (\hat{a}_M^\dagger)^{n_M}
                      |\mathrm{0}\rangle \nonumber \\
         =
            (-1)^{S_s} (\hat{a}_1^\dagger)^{n_1}
                       (\hat{a}_2^\dagger)^{n_2}    \cdots
                       (\hat{a}_s\hat{a}_s^\dagger) \cdots
                      (\hat{a}_M^\dagger)^{n_M}
                       |\mathrm{0}\rangle.
\end{align}

%
This phase factor complicates the application of
Hamiltonian~(Eq.~\ref{eq:ham_spinorb}) to the wavefunction.
Clearly, the operator $\hat{a}_s^{(\dagger)}$ acts beyond
its index $s$ and therefore the electronic SQR Hamiltonian,
in general, is not in the SOP form with respect
to the primitive degrees of freedom.
Wang and Thoss solved this issue by mapping the fermionic operators onto
equivalent spin operators~\cite{wan09:024114}.
Formally, this mapping consists in applying the inverse Jordan-Wigner (JW)
transformation to the fermionic field operators and effectively transforming the
fermionic Hamiltonian into an equivalent spin-chain Hamiltonian~\cite{jor28:631}.
The equivalent spin-$1/2$ chain Hamiltonian after the JW transformation reads~\cite{Sas20:154110}
\begin{align}
  \label{eq:ham_el_jw}
    \hat{H} = &
    \sum_{ij} h_{_{ij}}
        \left(
          \prod_{q=a+1}^{b-1} \hat{\sigma}^z_q
        \right) \hat{\sigma}_i^{+} \hat{\sigma}_j^{-} \nonumber\\
     + &  \frac{1}{2}\sum_{ijkl} v_{_{ijkl}}
          \left( \prod_{q=a+1}^{b-1} \hat{\sigma}^z_q
              \prod_{q^\prime=c+1}^{d-1} \hat{\sigma}^z_{q^\prime}
          \right) \nonumber\\
       &   \mathrm{sgn}(j-i)\mathrm{sgn}(l-k)
    \hat{\sigma}_i^{+} \hat{\sigma}_j^{+} \hat{\sigma}_l^{-} \hat{\sigma}_k^{-}.
\end{align}
where
$\hat{\sigma}_i^{+} = \frac{1}{2}\left( \hat{\sigma}_i^{x}+i\hat{\sigma}_i^{y} \right)$,
$\hat{\sigma}_i^{-} = \frac{1}{2}\left( \hat{\sigma}_i^{x}-i\hat{\sigma}_i^{y} \right)$, and
$\hat{\sigma}_k^{z}$ are the standard spin ladder operators with Pauli matrices
$\bm{\sigma}^{x}$, $\bm{\sigma}^{y}$ and $\bm{\sigma}^{z}$.
Here the indices $(a,b,c,d)$ correspond to the $(i,j,k,l)$ indices, but ordered from
smaller to larger and the $\mathrm{sgn}$ function is defined as
\begin{align}
    \mathrm{sgn}(x) =
        \begin{cases}
            1, & \text{if } x\geq 0\\
           -1, & \text{otherwise}.
        \end{cases}
\end{align}
The operators $\hat{\sigma}_\kappa^{+}$, $\hat{\sigma}_\kappa^{-}$
and $\hat{\sigma}^z_\kappa$ acts locally on $\kappa$-th spin-$\frac{1}{2}$
basis function and their matrix representation reads
\begin{align}
 \label{eq:mat_s_dof}
 \bm{\sigma}^{+} =
 \begin{pmatrix}
    0 & 0 \\
    1 & 0 \\
 \end{pmatrix} ;\,\,\,\,
 \bm{\sigma^{-}} =
 \begin{pmatrix}
    0 & 1 \\
    0 & 0 \\
 \end{pmatrix} ;\,\,\,\,
 \bm{\sigma^{z}} =
\begin{pmatrix}
    1 & 0 \\
    0 & -1 \\
\end{pmatrix} .
\end{align}
Clearly, the electronic Hamiltonian written in Eq.~\ref{eq:ham_el_jw} is in the
SOP form with respect to the primitive DOFs (spin-$\frac{1}{2}$ basis).
\begin{figure*}[t]
\begin{center}
\includegraphics[height=4.5cm]{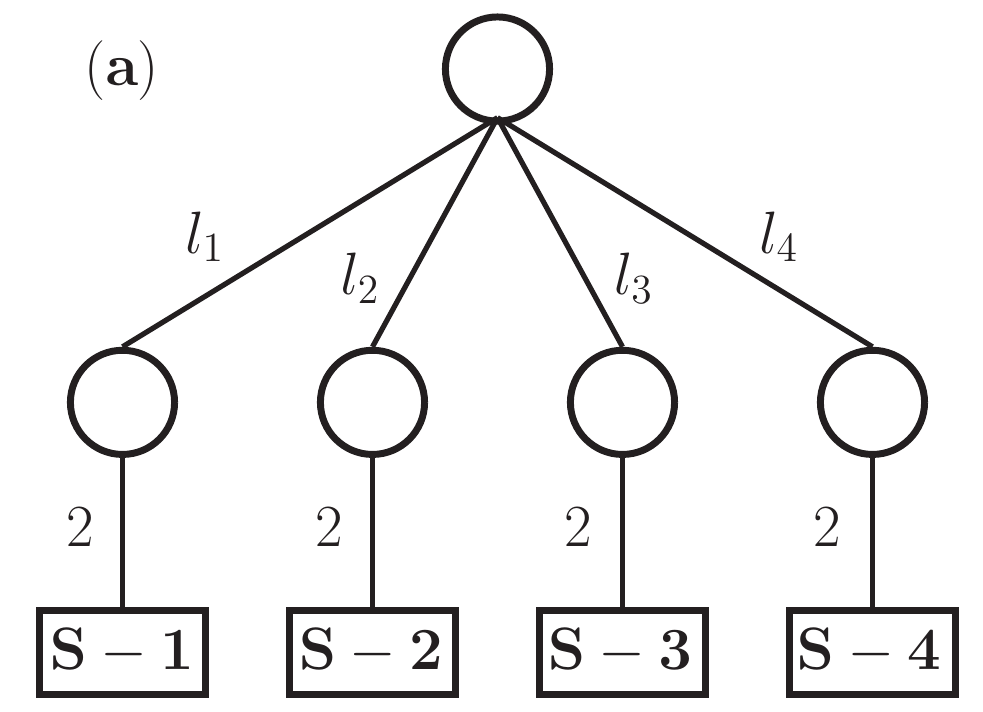}
\includegraphics[height=4.5cm]{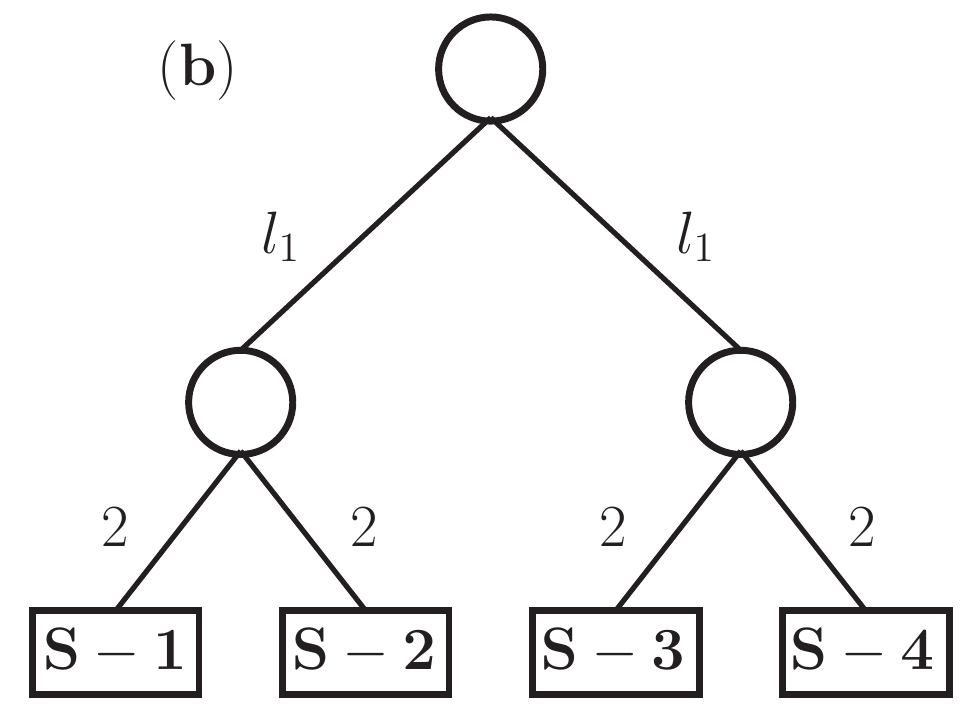}
\includegraphics[height=4.5cm]{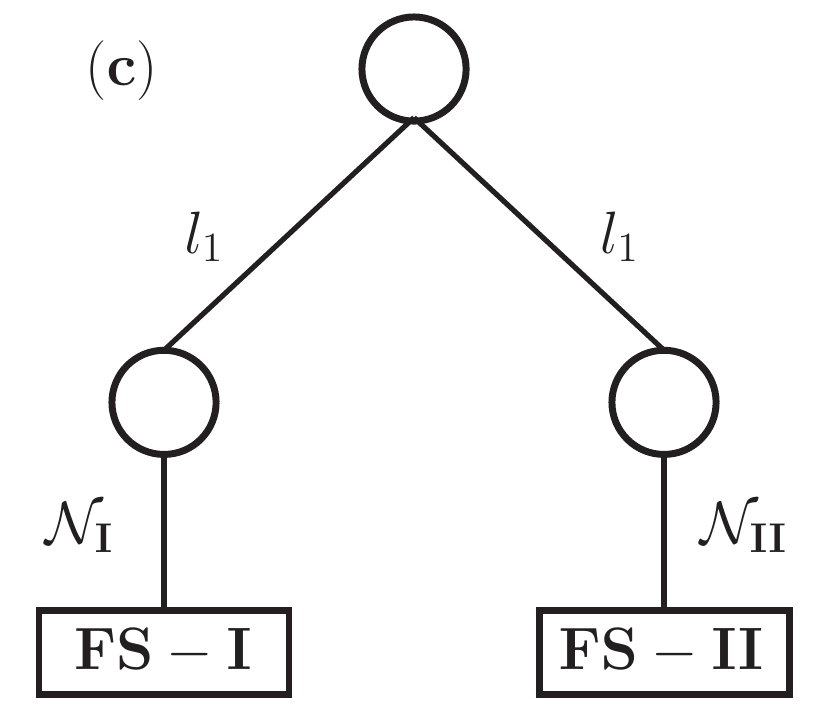}
\caption{Tree structures for the MCTDH-SQR wavefunction containing
         four spin orbitals:
         (a) Normal MCTDH wavefunction tree, in which spin degree of
             freedoms (S-DOF) are considered as primitive DOFs.
         (b) MCTDH wavefunction tree with mode combination, in which
             S-DOFs are considered as primitive DOFs.
         (c) MCTDH wavefunction tree, in which the Fock space of two spin
             orbitals is considered as single DOF (FS-DOF).
         $l_\kappa$ denotes number of SPF of the $\kappa$-th mode and
         $\mathcal{N}_{\kappa}$ is the number of configuration of
         $\kappa$-th FS-DOF.
    }
\label{fig:mctdhwfn}
\end{center}
\end{figure*}
\subsection{DOF in MCTDH-SQR: Spin and Fock space} \label{sec:theory:dof}
In the MCTDH-SQR formalism for fermionic system (without mode combination),
there are two limiting
wavefunction \emph{Ansätze}: Each spin degree of freedom (S-DOF) is
described by either (i) one or (ii) two time-dependent SPFs. The former corresponds to a
time dependent Hartree (TDH) wavefunction with a single
Hartree product and the latter corresponds to the exact wavefunction consisting of
2$^{M}$ configurations, where M is the number of S-DOF.
Thus, the limiting case (i) leads to a poor description of correlated state
and the later leads to an exact formulation, which becomes quickly unaffordable as the number
of S-DOF (i.e., spin orbitals) increases.
Therefore, the only practical
way to apply the MCTDH-SQR method is to create groups of S-DOFs, either
through mode combination or through its multi-layer generalization.
Figures~\ref{fig:mctdhwfn} (a) and (b) show the normal MCTDH wavefunction
and the MCTDH wavefunction with mode combination, respectively,
where S-DOFs are used as primitive DOFs.
\par
For $M$ spin orbitals there are 2$^M$ Fock states.
On the other hand one can divide the total Fock space into $f$ sub-Fock
spaces by grouping the M spin orbitals into $f$ groups
($m_1$, $m_2$, $\cdots$, $m_f$)
\begin{align}
 \mathcal{F}(M) = \mathcal{F}_1(m_1) \otimes \mathcal{F}_2(m_2)
 \otimes \dots \otimes \mathcal{F}_f(m_f)
 \label{eq:fsdof_fock}
\end{align}
where $\mathcal{F}_\kappa(m_\kappa)$ denotes the sub-Fock space of 
the $\kappa$-th FS-DOF (consists of $m_\kappa$ S-DOFs) with
2$^{m_\kappa}$ Fock states and
\begin{align}
 M = \sum_{\kappa=1}^{f} m_\kappa.
\end{align}
Now, one can represent the configurations of the sub-Fock space
$\mathcal{F}_\kappa(m_\kappa)$ as a new primitive DOF. We refer to this representation
of the primitive DOF as Fock space DOF (FS-DOF).
Fig.~\ref{fig:mctdhwfn} (c) shows the MCTDH wavefunction involving FS-DOF
as primitive DOF.
In the FS-DOF formalism, one needs to transform the primitive operator string described in
the Eq.~\ref{eq:ham_el_jw} into new primitive matrix operators acting onto the sub-Fock
spaces of their corresponding FS-DOF.

In the FS-DOF representation, the state of the system is described by \emph{kets} $|i_1,\ldots,i_f\rangle$, where
$i_\kappa$ corresponds to the $i$-th configuration of the $\kappa$-th FS-DOF. One can think of the
$i_\kappa$ configuration indices as indexing Euclidean basis vectors within each degree of freedom.
The configurations
within a FS-DOF can correspond, in general, to different electron occupation numbers.
A matrix element of the total Hamiltonian is then represented as
\begin{align}
    \label{eq:ham_fsdof}
    H_{z_1,\ldots,z_f} = H_{i_1,\ldots,i_f}^{j_1,\ldots,j_f} =
    \langle j_1,\ldots,j_f| \hat{H} | i_1,\ldots,i_f \rangle,
\end{align}
where $z_\kappa = (i_\kappa, j_\kappa)$ are double indices indicating
the \emph{bra}- and \emph{ket}-side configurations within each FS-DOF.
This electronic Hamiltonian is very sparse and, clearly, it can only be explicitly
constructed and diagonalized for small systems. In the following, thus, we will discuss
the construction of sum-of-product (SOP) forms of Hamiltonian~(\ref{eq:ham_fsdof}), either
exact or approximate, and their
use within the framework of MCTDH-SQR.

\subsection{Compact form of the Hamiltonian in the FS-DOF basis} \label{sec:theory:compact}
The number of terms of the electronic SQR Hamiltonian
(Eqs.~\ref{eq:ham_spinorb},~\ref{eq:ham_el_jw}) increases with M$^4$ where
M is the total number of spin orbitals under consideration.
One can only reduce the number of terms by applying
cutoffs to different types of one- and two-electron integrals at the expense
reducing the accuracy of the Hamiltonian~\cite{Sas20:154110}.
In the following, we formulate two strategies
to reduce the number of Hamiltonian terms in the SOP form of the Hamiltonian.
Both strategies are based on the formation of FS-DOF as the primitive DOF and
therefore in both strategies the MCTDH wavefunction structure remains the same.
The difference is in the way the Hamiltonian is
constructed, either (i) directly starting from the original SQR Hamiltonian and
mapping each product term (each chain of spin operators) to its matrix representation in the FS-DOF basis
or (ii) constructing the matrix elements of the product operator directly as an optimal
SOP fit to the full operator.

\subsubsection{Summed SQR Hamiltonian: S-SQR}
\label{sec:theory:spinorb}
As a starter, the number of terms of
the SQR Hamiltonian in its spin-$\frac{1}{2}$ (S-DOF)
form, Eq.~\ref{eq:ham_el_jw}, can be reduced by summing up those two-body terms that can be
transformed into the same order of the ladder operators. In other words, one
should use the symmetry of the four-index tensor $v_{ijkl}$ to avoid unnecessary
work when applying the operator.
The 2-index correlated operators that can be added together are
\begin{align}
 \label{eq:sdof_2i}
   v_{ijij} \hat{a}_i^{\dagger} \hat{a}_j^{\dagger} \hat{a}_j \hat{a}_i
 + v_{ijji} \hat{a}_i^{\dagger} \hat{a}_j^{\dagger} \hat{a}_i \hat{a}_j
 = (v_{ijij}-v_{ijji}) \hat{a}_i^{\dagger} \hat{a}_j^{\dagger} \hat{a}_j \hat{a}_i .
\end{align}
The 3-index correlated operators that can be added together are
\begin{align}
 \label{eq:sdof_3i}
 &  v_{ijik} \hat{a}_i^{\dagger} \hat{a}_j^{\dagger} \hat{a}_k \hat{a}_i
 +  v_{ijki} \hat{a}_i^{\dagger} \hat{a}_j^{\dagger} \hat{a}_i \hat{a}_k
 + \nonumber \\
 &  v_{jiki} \hat{a}_j^{\dagger} \hat{a}_i^{\dagger} \hat{a}_i \hat{a}_k
 +  v_{jiik} \hat{a}_j^{\dagger} \hat{a}_i^{\dagger} \hat{a}_k \hat{a}_i \nonumber \\
 &= (v_{ijik} - v_{ijki} + v_{jiki} - v_{jiik})
        \hat{a}_i^{\dagger} \hat{a}_j^{\dagger} \hat{a}_k \hat{a}_i .
\end{align}
The 4-index correlated operators that can be added together are
\begin{align}
 \label{eq:sdof_4i}
 &  v_{ijkl} \hat{a}_i^{\dagger} \hat{a}_j^{\dagger} \hat{a}_l \hat{a}_k
 +  v_{ijlk} \hat{a}_i^{\dagger} \hat{a}_j^{\dagger} \hat{a}_k \hat{a}_l
 + \nonumber \\
 &  v_{jilk} \hat{a}_j^{\dagger} \hat{a}_i^{\dagger} \hat{a}_k \hat{a}_l
 +  v_{jikl} \hat{a}_j^{\dagger} \hat{a}_i^{\dagger} \hat{a}_l \hat{a}_k \nonumber \\
 &= (v_{ijkl} - v_{ijlk} + v_{jilk} - v_{jikl})
    \hat{a}_i^{\dagger} \hat{a}_j^{\dagger} \hat{a}_l \hat{a}_k .
\end{align}
However, the reduction in the number of terms achieved by the sums
in Eqs.~\ref{eq:sdof_2i}, \ref{eq:sdof_3i}
and \ref{eq:sdof_4i} is negligible compared to the number of terms in the SQR
Hamiltonian.
The situation improves substantially while
using FS-DOF as primitive DOF. In the FS-DOF primitive basis, all terms
(both one- and two-body) that act only within the spin-orbitals of one FS-DOF can be summed up
to form an uncorrelated operator term. Similarly, one can form correlated operators (products acting on two or more FS-DOF)
by summing all terms that act within one FS-DOF for each
distinct string operating
on the spin orbitals of the other FS-DOF.
\par
For example, for the FS-DOF construction described in Fig.~\ref{fig:mctdhwfn},
one can form the uncorrelated operator of the 1st FS-DOF (FS-I) by summing up the
following terms
\begin{align}
 &  h_{11}(\hat{a}_1^{\dagger} \hat{a}_1)^{(I)} + h_{12}(\hat{a}_1^{\dagger} \hat{a}_2)^{(I)}
  + h_{21}(\hat{a}_2^{\dagger} \hat{a}_1)^{(I)} + h_{22}(\hat{a}_2^{\dagger} \hat{a}_2)^{(I)} \nonumber\\
 &+ v_{1212}(\hat{a}_1^{\dagger} \hat{a}_2^{\dagger} \hat{a}_2 \hat{a}_1)^{(I)}
  + v_{1221}(\hat{a}_1^{\dagger} \hat{a}_2^{\dagger} \hat{a}_1 \hat{a}_2)^{(I)} .
\end{align}
Of course, one needs to transform the fermionic operator strings
into spin operators using the JW transformation and then form the matrix operator
of the corresponding chain of operators in the space of the FS-DOF
before the summation operation.
One of the correlated operator terms can be formed by summing up the following contributions
\begin{align}
 \label{eq:corr_fsdof}
 h_{13} \hat{a}_1^{\dagger} \hat{a}_3 + h_{23} \hat{a}_2^{\dagger} \hat{a}_3
 &= h_{13} \hat{\sigma}_1^{+} \hat{\sigma}^z_2 \hat{\sigma}_3^{-}
  + h_{23} \hat{\sigma}_2^{+} \hat{\sigma}_3^{-} \nonumber\\
 &= \left(
          h_{13}(\hat{\sigma}_1^{+} \hat{\sigma}^z_2)^{(I)}+h_{23}(\hat{\sigma}_2^{+})^{(I)}
    \right)
    (\hat{\sigma}_3^{-})^{(II)} .
\end{align}
%
Please note that there are other two-body terms that can also be added to the terms described
in Eq.~\ref{eq:corr_fsdof}, for example, $v_{1213}$, $v_{1231}$, $v_{2131}$, $v_{2113}$,
$v_{2123}$, $v_{2132}$, $v_{1232}$, and  $v_{1223}$ terms associated with their corresponding
operator string.
The compact form of SQR Hamiltonian for an arbitrary FS-DOF combination is given
in Appendix~\ref{ap:ham_spinorb}.

The S-SQR form of the Hamiltonian is exact within the space of configurations spanned
by the different FS-DOF.
Since the matrix operators acting on each FS-DOF and for different products
are not related to each other,
this form
of the SQR operator corresponds to an exact canonical polyadic decomposition
(CANDECOMP)~\cite{Hit27:164, har70:1, Car70:238, Kie00:105, Sch20:024108}
In general, it reads
\begin{align}
    \label{eq:candecomp}
    H_{z_1,\ldots,z_f} = \sum_{s=1}^R \prod_{\kappa=1}^f [\mathbf{X}_s^{(\kappa)}]_{z_\kappa},
\end{align}
where any multiplicative constants are absorbed into the matrices.
The final rank $R$, and hence the degree of compactification of the Hamiltonian, depends on the grouping of
the orbitals to form the FS-DOF. In the limiting case that all spin orbitals are
grouped together to form just one FS-DOF, $R=1$. This is not very useful as the size
of the only operator matrix is $2^M\times 2^M$
and corresponds to the full configuration interaction Hamiltonian in Fock space.
In the case that no grouping is performed (e.g. in a ML-MCTDH calculation without mode-combination)
the Hamiltonian remains identical with the original SQR Hamiltonian and no gain through summation
is achieved.
%

\subsubsection{Tucker decomposition of the SQR Hamiltonian: T-SQR}
\label{sec:theory:tucker}
An alternative approach to a SOP form
abandons the exact representation of the Hamiltonian and introduces an optimal
Tucker decomposition of the electronic SQR Hamiltonian $H_{z_1,\ldots,z_f}$
in the FS-DOF primitive basis,
\begin{align}
 H_{z_1,\ldots,z_f}^{(T)} \approx \sum_{l_1}^{n_1} \cdots \sum_{l_f}^{n_f}
                g_{l_1,\cdots,l_f}
                \prod_{\kappa=1}^{f} [\mathbf{O}^{(\kappa,l_\kappa)}]_{z_\kappa},
 \label{eq:elham_tucker}
\end{align}
where $g_{l_1,\cdots,l_f}$ are the elements of
the Tucker core tensor with rank $(n_1, n_2, \cdots, n_f)$ and
$\mathbf{O}^{(\kappa,l_{\kappa})}$ is the $l_{\kappa}$-th single particle operator (SPO)
acting of the $\kappa$-th FS-DOF.
The operator matrices $\mathbf{O}^{(\kappa,l_{\kappa})}$ are defined in the
sub-Fock space ($\mathcal{F}_\kappa(m_\kappa)$) formed by the $\kappa$-th
FS-DOF.
When multiplied as vectors, these elements form an orthonormal set,
\begin{align}
    \label{eq:O_ortho}
    \sum_{z_\kappa} [\mathbf{O}^{(\kappa, l_\kappa)}]_{z\kappa} \cdot [\mathbf{O}^{(\kappa, p_\kappa)}]_{z\kappa} = \delta_{l_\kappa, p_\kappa}.
\end{align}


%
The challenge here is to obtain the core tensor $g$ and the basis of SPO matrices.
Similarly to the original \texttt{POTFIT}~\cite{bec00:1,jaec96:7974,jaec98:3772}
algorithm for multidimensional potential energy surfaces, this can be formulated as the numerical
problem of finding optimal coefficients with respect to the minimization of
some error function. One of such error function is the sum of squared errors
 \begin{align}
 \label{eq:tuckerr}
     \mathcal{L}^2 = 
     \sum_{z_1} \cdots \sum_{z_f}
     \Bigg( &
           H_{z_1,\ldots,z_f} - H_{z_1,\ldots,z_f}^{(T)}
     \Bigg) ^2
 \end{align}
where $H_{z_1,\ldots,z_f}^{(T)}$ takes the form of Eq.~(\ref{eq:elham_tucker}).

%
The Hamiltonian tensor elements $H_{z_1,\ldots,z_f}$ entering Eq.~(\ref{eq:tuckerr}) can be evaluated
either directly using the Slater-Condon rules in the full configuration space~\cite{Sla29:1293,Con30:1121},
or by using the representation of Eq.~(\ref{eq:candecomp}).
The rank of the original Hamiltonian tensor is
$\prod_{\kappa} \mathcal{N}_\kappa^2 = 2^{2M}$.
Since, the Tucker rank ($\prod_\kappa n_\kappa$) is,
in general, much smaller than the rank of the original Hamiltonian,
the Tucker decomposed Hamiltonian can easily achieve a more compact form compared to the
original Hamiltonian tensor.
Nonetheless, the rank of the SQR Hamiltonian in the
full Fock space scales much more rapidly than $M^4$, although its matrix representation
is extremely sparse.
It is therefore yet to be tested numerically whether on optimal Tucker representation of the operator
can bring an advantage compared to the previously introduced, exact summation strategy.
%

The rank of the Hamiltonian in Eq.~\ref{eq:elham_tucker} in the
SOP form, $\prod_\kappa n_\kappa$
can still be reduced by a factor of $n_\nu$, where $\nu$ is the index of any of the
FS-DOFs, by contracting this degree of freedom with the core tensor and
summing up all operators of the $\nu$ FS-DOF that multiply a common string of
operators in the other FS-DOFs. This operation is standardly performed in the POTFIT algorithm for
potential operators.

In practice, we have used the \texttt{TensorLy}~\cite{tensorly}, python library
that uses \textit{higher-order orthogonal iteration} (HOOI)~\cite{kol09:455}
to obtain the Tucker decomposition of the electronic Hamiltonian.
As in the original potfit algorithm, the full tensor needs to be stored in memory,
which prevents the application of this approach
to large molecular systems where the original tensor cannot be stored in
memory.
We are not concerned with this limitation for
the proof-of-concept application to molecular electronic
dynamics, and will address it in future
work resorting, e.g., to sampling strategies
over the elements of the primitive tensor~\cite{Sch17:064105,Sch20:024108}.

Finally, we note that the Tucker decomposed Hamiltonian obtained
by numerically minimizing the error function (Eq.~\ref{eq:tuckerr})
is not completely Hermitian, as Hemiticity of the product terms is not explicitly enforced in the
fitting form.
This results in small numerical inaccuracies related to norm conservation of the wavefunction.
Hermiticity can be restored, e.g., if the space of SPOs within each FS-DOF
is spanned by Hermitian matrices $\mathbf{A}^{(\kappa, j_\kappa)} = {\mathbf{A}^\dagger}^{(\kappa, j_\kappa)}$
and by pairs of Hermitian conjugate matrices $\mathbf{B}^{(\kappa, j_\kappa)}$,
$\mathbf{B}{^\dagger}^{(\kappa, j_\kappa)}$.
Non-Hermitian SPOs are crucial for representing the electronic dynamics across
different FS-DOF (in a similar way as the original creation and annihilation operators) but each
term containing $\mathbf{B}^{(\kappa, j_\kappa)}$ must be matched by a corresponding term with the same
core-tensor coefficient and containing $\mathbf{B}{^\dagger}^{(\kappa, j_\kappa)}$.
Instead of enforcing these constraints during the minimization of $\mathcal{L}^2$, which
is more efficient,
currently we opted
for reintroducing Hermiticity \emph{a posteriori}.
To this end we average each product term in
Eq~\ref{eq:elham_tucker} with its Hermitian conjugate, thus
turning the Hamiltonian exactly Hermitian
with a factor 2 more terms.
\subsubsection{Pruning of the FS-DOF sub-Fock space}  \label{sec:theory:pruning}
As a final remark, grouping the primitive S-DOF as FS-DOF allows for each sub-Fock space to be statically pruned by restricting the number of
electrons and/or by removing configurations with unwanted orbital
occupation.
This is equivalent of removing undesired grid points from a multidimensional
grid in a discrete variable representation.
As an example, let us consider a system containing two FS-DOF each
containing four spin orbitals.
The number of configurations ($\mathcal{N}$) in each FS-DOF is $2^4=16$.
Let's also consider the system contains four electrons and the orbitals
are energetically ordered with respect to a pre-existing first quantization
mean-fields calculation. So, the lowest energy configuration is where all
the orbitals are occupied ($|1,1,1,1\rangle$) in the 1st FS-DOF
and all the orbitals in the 2nd FS-DOF are empty ($|0,0,0,0\rangle$).
Now, if one allows only up to two holes in the 1st FS-DOF and
two particles in the 2nd FS-DOF, the number of configurations in each
FS-DOF shrinks from 16 to 11. One can further removed the configuration
of the 1st FS-DOF where both of the lowest two spin orbitals are empty
by removing ($|0,0,1,1\rangle$) if chemically irrelevant and shrinks
the number of configurations ($\mathcal{N}_1$) to 10.
Similarly, if chemically irrelevant, one can further remove the
$|0,0,1,1\rangle$ occupation number state from the 2nd FS-DOF
to shrink the number of configurations ($\mathcal{N}_2$) to 10.
The pruning of the FS-DOF allows one to reduce the primitive space
of the FS-DOF without compensating too much chemical accuracy.
Another advantage of the (pruned) FS-DOF representation is that the SQR operators acting on the SPFs of the corresponding FS-DOF are very sparse and
correspond to mappings (see Appendix~\ref{ap:matvecmul}) that can lead to large efficiency gain in computations.
\par
\section{Results and Discussion}           \label{sec:results}
\subsection{Electronic eigenenergies of LiH}  \label{sec:res:lih}
We have calculated the adiabatic electronic energies of LiH by
applying the MCTDH-SQR method to the electronic SQR Hamiltonian
at fixed nuclear geometries. The 6-31G atomic basis is used for
both Li and H. In first quantization, the potential
energy curves (PECs) of the lowest four electronic states are obtained
for comparison at the full configuration interaction level.
In SQR, the PECs are obtained through the propagation of an MCTDH-SQR
wavefunction which overlaps with the various excited electronic states.
The initial condition for the singlet states is
\begin{align}
 \label{eq:lih_initwf_singlet}
   |\Psi(t=0)\rangle
    &= | 1_\alpha, 1_\beta, 2_\alpha, 2_\beta  \rangle \nonumber\\
    &+ | 1_\alpha, 1_\beta, 2_\alpha, 3_\beta  \rangle
     - | 1_\alpha, 1_\beta, 2_\beta, 3_\alpha  \rangle ,
\end{align}
and for the triplet states is
\begin{align}
 \label{eq:lih_initwf_triplet}
   |\Psi(t=0)\rangle
    &= | 1_\alpha, 1_\beta, 2_\alpha, 3_\beta  \rangle
     + | 1_\alpha, 1_\beta, 2_\beta, 3_\alpha  \rangle \nonumber\\
    &+ | 1_\alpha, 1_\beta, 2_\alpha, 6_\beta  \rangle
     + | 1_\alpha, 1_\beta, 2_\beta, 6_\alpha  \rangle .
\end{align}
The wavefunction in Eqs.~(\ref{eq:lih_initwf_singlet}) and (\ref{eq:lih_initwf_triplet})
is spin-singlet and spin-triplet and overlaps with the desired $^1\Sigma^{+}$ and
$^3\Sigma^{+}$ states, respectively.
The electronic eigenenergies are obtained from the maxima of the peaks in
the power spectrum obtained from the Fourier transform of the
autocorrelation function
\begin{align}
 \sigma(E) = \frac{1}{\pi}Re\int_{0}^{\infty} e^{iEt}
 \langle \Psi|\Psi(t) \rangle dt .
 \label{eq:power_spectrum}
\end{align}
\begin{table}[t]
 \caption{Number of Hamiltonian terms,  memory required to store the operator
          matrices and wall-clock time for the different representation of
          Hamiltonian. The wall-clock time is given for 1 fs time propagation.
          The calculations have been performed with 14 CPUs using shared-memory
          parallelization on the same machine and CPU type,  namely, Dual-Core
          Intel Xeon,  processor type E5-2650 v2 running at 2.6 GHz and the
          wall-clock times are intended for their relative comparison only.
          Matrix and mapping correspond to usual
          and mapping strategy
          (see Appendix~\ref{ap:matvecmul}, Algorithm~\ref{al:mapping})
          used for the matrix vector multiplication, respectively.
          Maximum rank of the T-SQR Hamiltonian is (6241, 6241)
          }
 \begin{ruledtabular}
  {\begin{center}
    \begin{tabular}{lcccr}
     Method & Hamil. terms & Size &Time (h:m)\\
     \hline
     {\bf SQR} \\
     mapping & 16138 & 465 KB & 7:14 \\
     \hline
     {\bf S-SQR} \\
     matrix    & 2003  & 394 MB & 1:43 \\
     mapping   & 2003  & 5 MB   & 1:19 \\
     \hline
     {\bf T-SQR} \\
     (20, 20)   & 40  & 703 KB & 0:01 \\
     (40, 40)   & 80  & 1.3 MB & 0:01 \\
     (60, 60)   & 120 & 2.5 MB & 0:01 \\
     (80, 80)   & 160 & 3.1 MB & 0:02 \\
     (100, 100) & 200 & 3.6 MB & 0.02 \\
     (150, 150) & 300 & 4.7 MB & 0.03
    \end{tabular}
   \end{center} }
  \end{ruledtabular}
 \label{tab:lih_pes} 
\end{table}
\begin{figure}[t]
 \begin{center}
  \includegraphics[width=8.5cm]{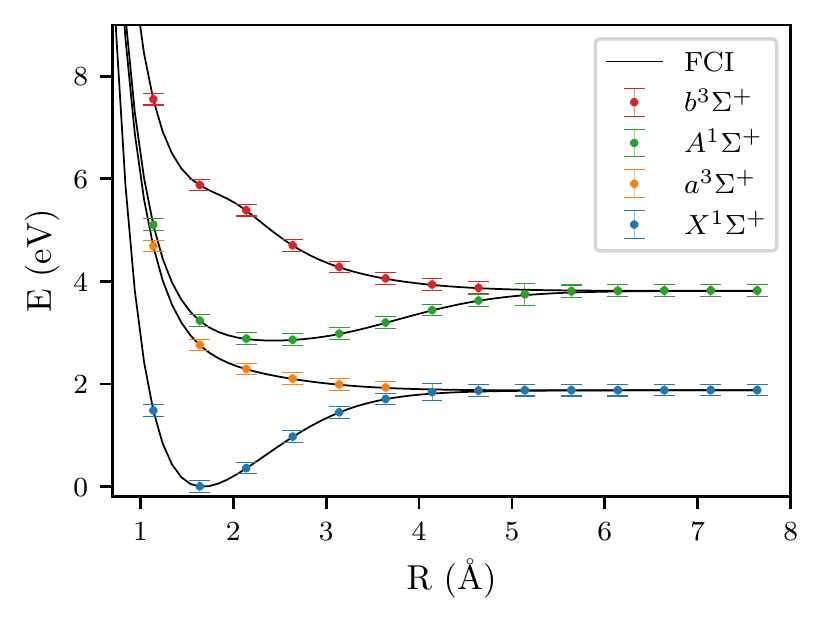}
  \caption{Comparison of eigenenergies of S-SQR Hamiltonian with FCI.}
  \label{fig:lih_pes_spinorb}
 \end{center}
\end{figure}
\begin{figure*}[t]
 \begin{center}
  \includegraphics[width=5.6cm]{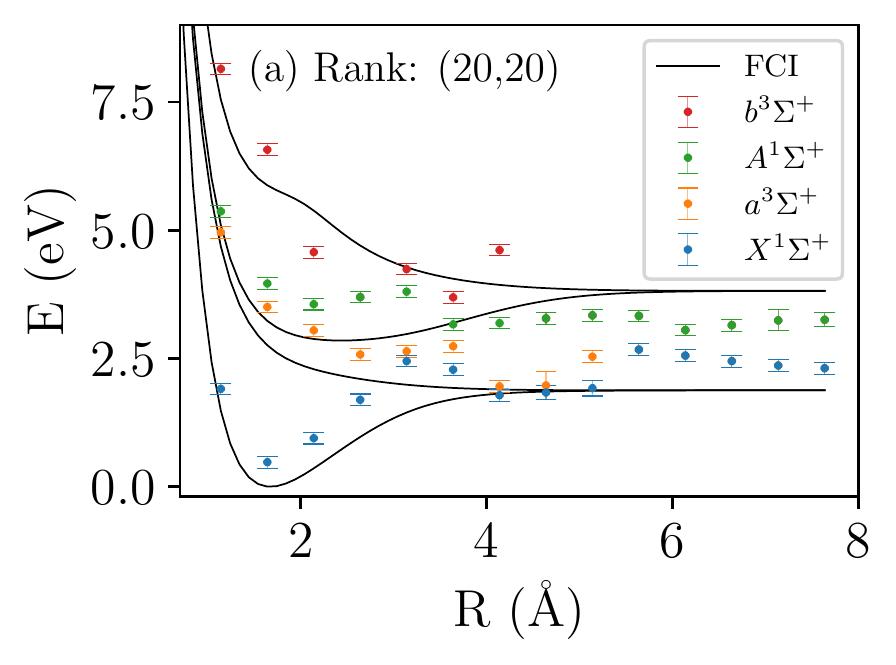}
  \includegraphics[width=5.6cm]{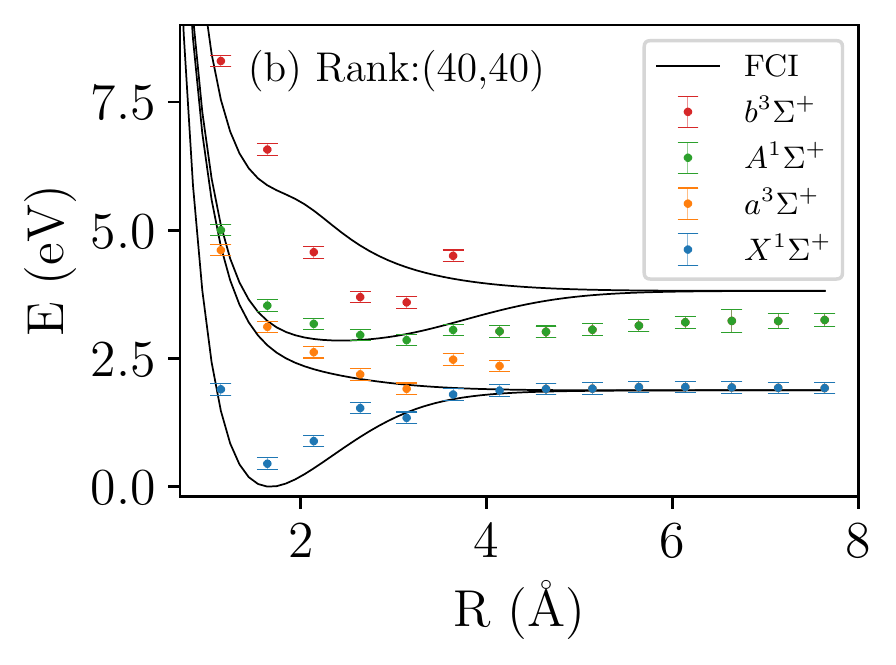}
  \includegraphics[width=5.6cm]{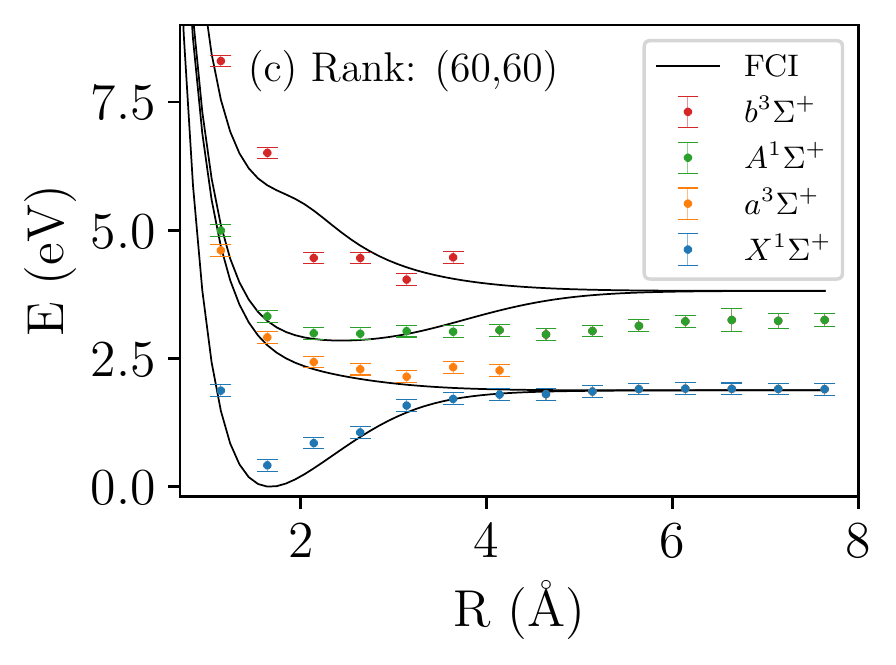}
  \includegraphics[width=5.6cm]{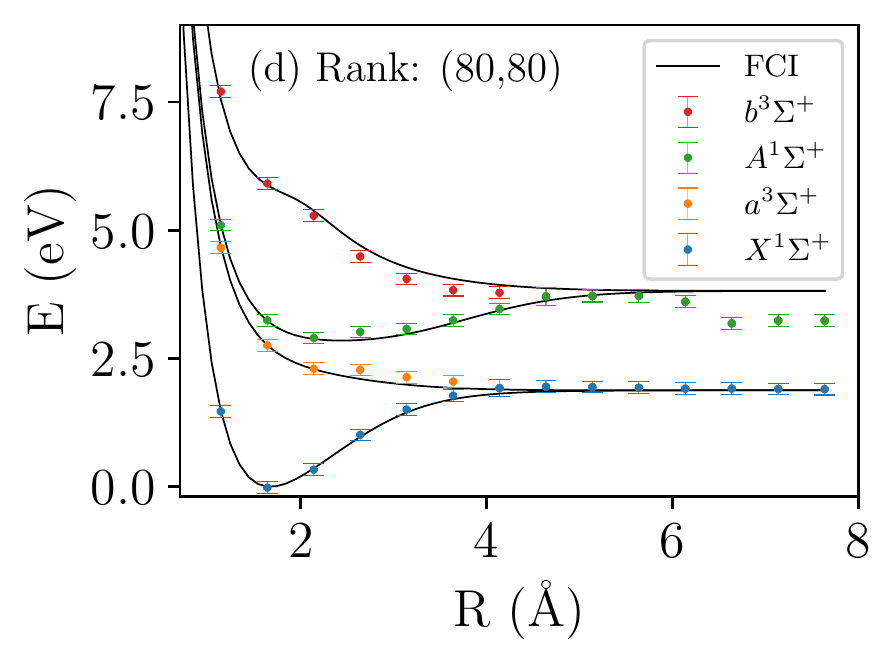}
  \includegraphics[width=5.6cm]{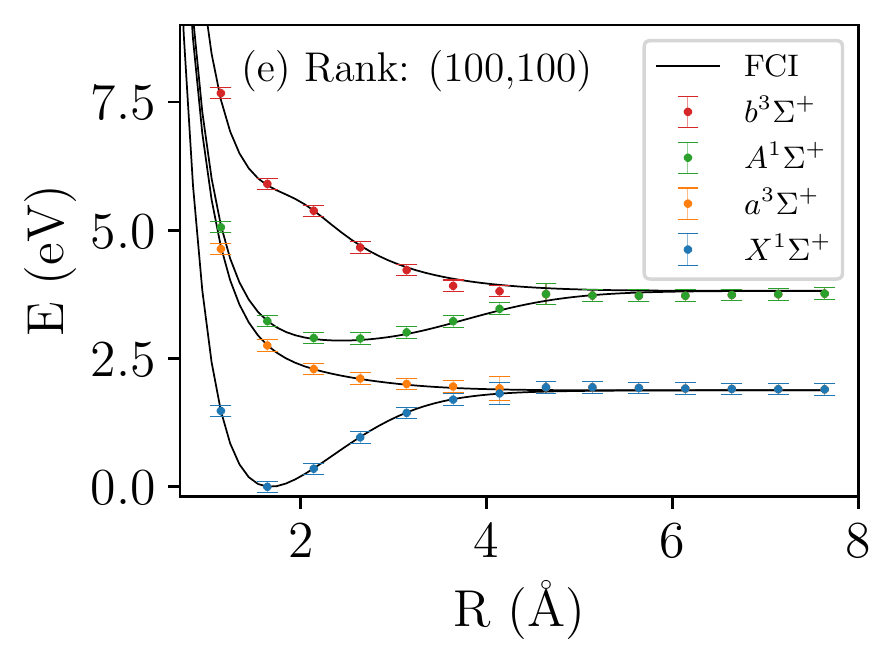}
  \includegraphics[width=5.6cm]{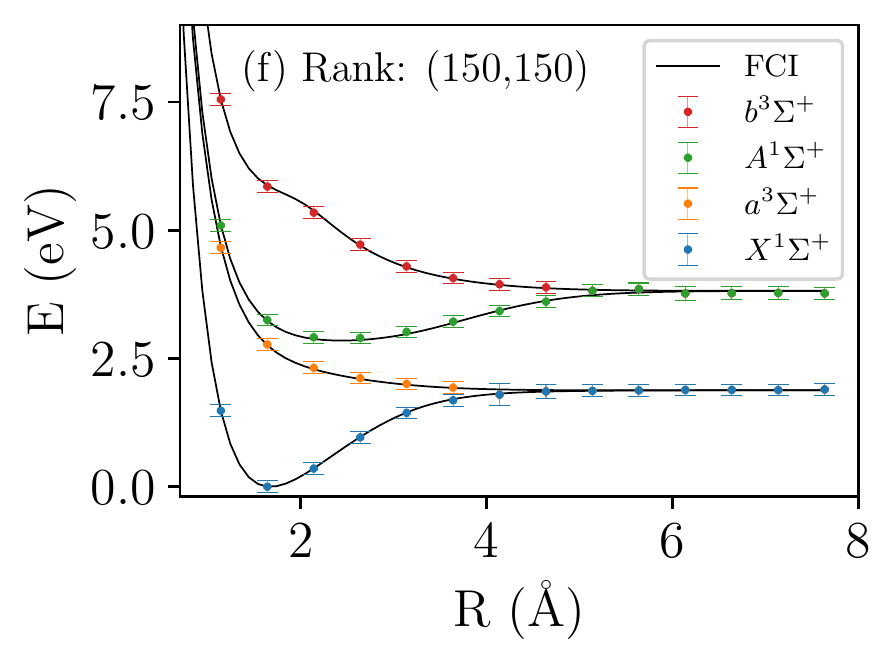}
  \caption{Comparison of electronic eigenenergies in the
           T-SQR Hamiltonian approach.
           Electronic eigenenergies calculated with Tucker rank
           (20,20), (40,40), (60,60), (80,80), (100,100) and (150,150)
           are shown in (a), (b), (c), (d), (e) and (f), respectively.
           Maximum rank of the T-SQR Hamiltonian is (6241,6241).
           Black lines are the potential energy curves calculated in the FCI method.
           }
  \label{fig:lih_pes_tucker}
 \end{center}
\end{figure*}
\par
The FS-DOFs for the SQR calculations are grouped in the following way.
The lowest 5 spatial (10 spin) molecular orbitals are arranged into
the first FS-DOF and the
remaining 6 spatial (12 spin) molecular orbitals into the second FS-DOF.
This results in 2$^{10}$ and 2$^{12}$ configurations in the first and
second FS-DOF, respectively. These Fock subspaces are then pruned
using the following procedure. (i) In the first FS-DOF we allow (0$-$2) alpha,
(0$-$2) beta and (2$-$4) total electrons. We further prune this FS-DOF
by removing all configurations where the lowest energy MOs (1s of Li)
are completely empty (i.e. both alpha and beta electrons missing form
this spatial orbital). This shrinks the number of configurations
of this FS-DOF from 1024 to 133. (ii) Similarly in the second
FS-DOF, we allow (0$-$2) alpha, (0$-$2) beta and (0$-$2) total electrons.
This shrinks the number of configurations of this FS-DOF from
4096 to 79.
\par
The electronic SQR Hamiltonian is represented either in S-SQR or T-SQR forms.
Table \ref{tab:lih_pes} compares the number of Hamiltonian terms, memory
required to store the operator matrices and CPU time for different calculations
in each of the two approaches.
For S-SQR Hamiltonian, we have compare calculations where the
matrices of each Hamiltonian terms are stored
(and thus, applied to the wavefunction) as a matrix or
as a mapping (see Appendix~\ref{ap:matvecmul}).
For the mapping case, we consider in turn two situations $-$
(i) the original SQR Hamiltonian is applied without summing the uncorrelated and correlated terms,
(ii) the Hamiltonian terms are compacted as much as possible to form S-SQR Hamiltonian and then applied to
the wavefunction. For the operator in matrix form, the
MCTDH code~\cite{mctdh:MLpackage}
by default sums up the Hamiltonian terms.
From the Table \ref{tab:lih_pes},
it is clear that the uncompacted (mapping) form of the Hamiltonian is the
slowest. On the other hand, the compacted S-SQR Hamiltonian using the mapping
for the matrix-vector multiplications is fastest and requires less
memory compared to the S-SQR Hamiltonian matrix form.
Note that with identical MCTDH parameters, all of the
three calculations yield identical results within numerical accuracy.
The electronic eigenenergies in the spin orbital basis along with the PECs
obtained from the FCI calculation are shown in Fig. \ref{fig:lih_pes_spinorb}.
\par
\begin{figure}[t]
 \begin{center}
  \includegraphics[width=4.25cm]{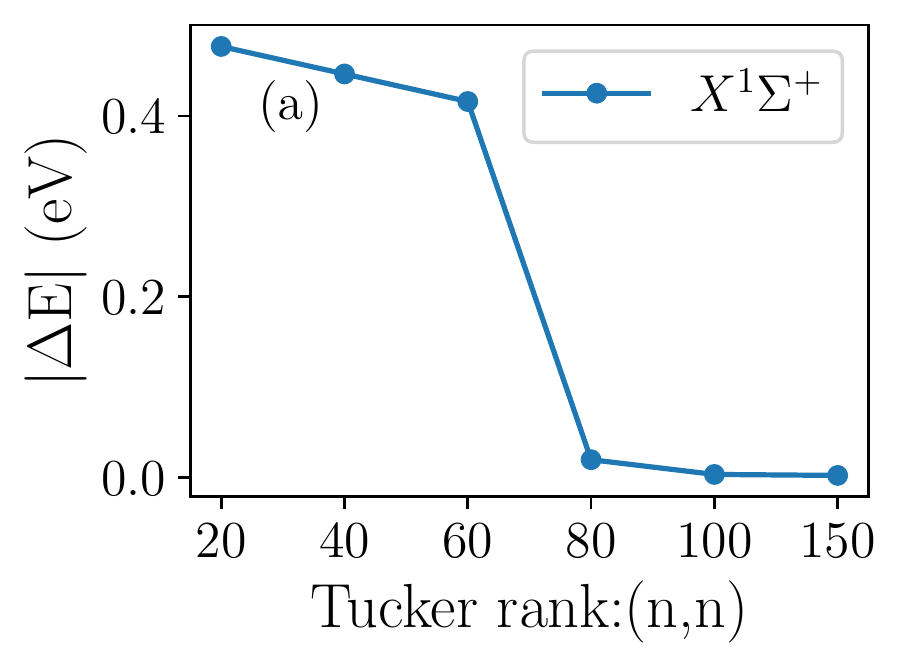}
  \includegraphics[width=4.25cm]{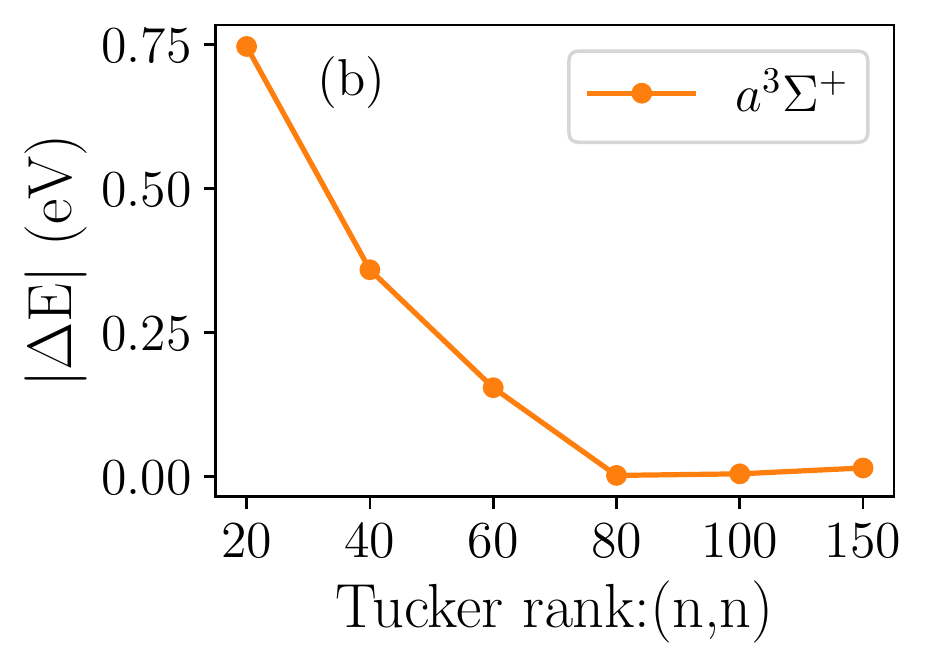}
  \includegraphics[width=4.25cm]{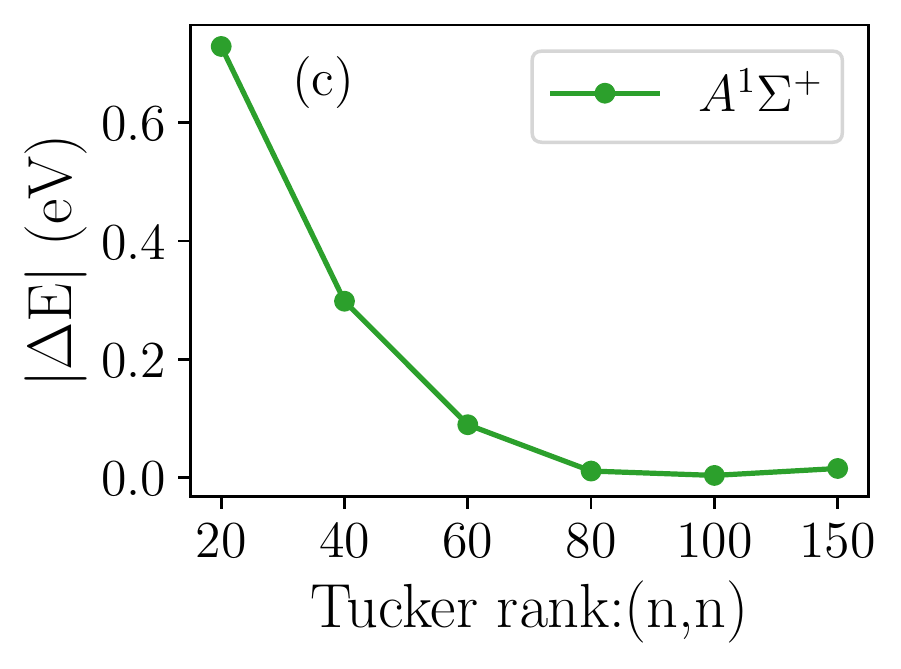}
  \includegraphics[width=4.25cm]{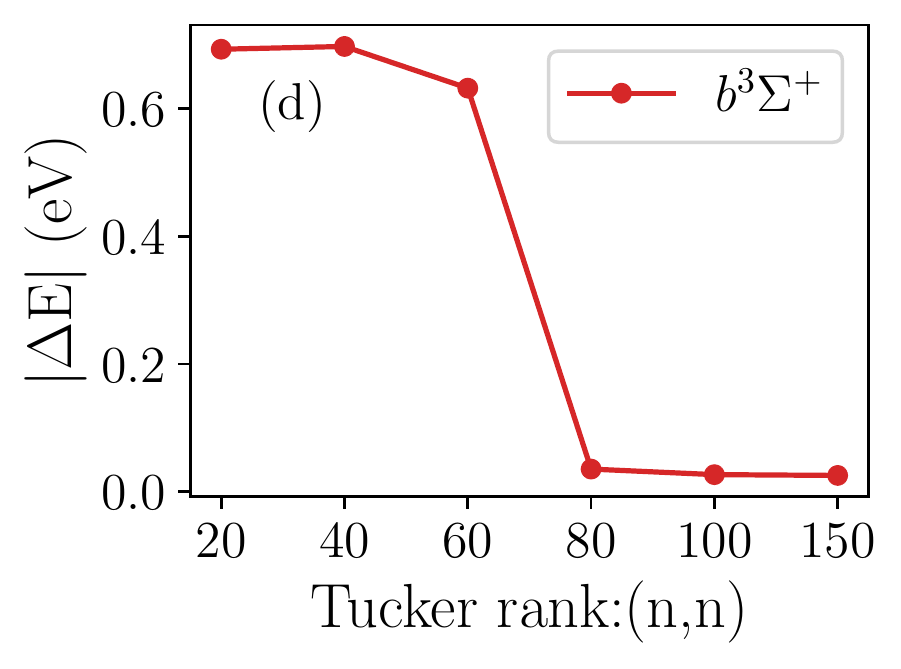}
  \caption{Convergence of electronic eigenenergy (at 1.64 au bond length) with
           different Tucker rank.
           Difference of calculated electronic energy from the FCI energy
           for the four electronic states (a) $X^{1}\Sigma^{+}$, (b) $a^{3}\Sigma^{+}$,
           (c) $A^{1}\Sigma^{+}$ and (d) $b^{3}\Sigma^{+}$ is shown.
           }
\label{fig:lih_tuck_conv}
\end{center}
\end{figure}

Next, we compare the electronic Hamiltonian in the T-SQR Hamiltonian approach.
In Table \ref{tab:lih_pes}, the number of Hamiltonian terms, memory
required to store the operator matrices and CPU time are compared as a
function of the rank of the Tucker decomposition.
The calculated eigenenergies for different Tucker rank are shown in
Fig. \ref{fig:lih_pes_tucker}. As expected, the electronic energies are
gradually improving with increasing Tucker rank. This is illustrated
in Fig. \ref{fig:lih_tuck_conv}, where we show the convergence of the eigenenergies
($\Delta E = E_{FCI}-E_{T\mbox{-}SQR}$) at the equilibrium geometry (at $R=1.64~$\AA)
with respect to the Tucker rank.
A very good convergence is achieved with a Tucker rank (100,100), much smaller than
the full rank (6241, 6241).
For this small example using the T-SQR Hamiltonian approach, the execution time
for the converged
calculation of comparable accuracy is $\sim$ 26 times faster
(cf. Table~\ref{tab:lih_pes}) than the best possible compact form of
the S-SQR Hamiltonian approach. This is in part achieved because
of the contraction of the operator for one of the two FS-DOFs.

\subsection{Electronic ionization spectrum of H$_2$O}  \label{sec:res:h2o}
The \emph{ab initio} MCTDH-SQR approach can be directly
applied to the calculation of ionization spectra of molecular systems through
time-propagation instead of matrix diagonalization. Here we benchmark the method on the water molecule.
The 6-31G atomic basis is used for both O and H to generate the MOs of H$_2$O.
The lowest energy MO (1s of O) is frozen and the remaining 12 spatial orbitals are
considered for the calculation.
\begin{figure*}[t]
 \begin{center}
  \includegraphics[width=5.6cm]{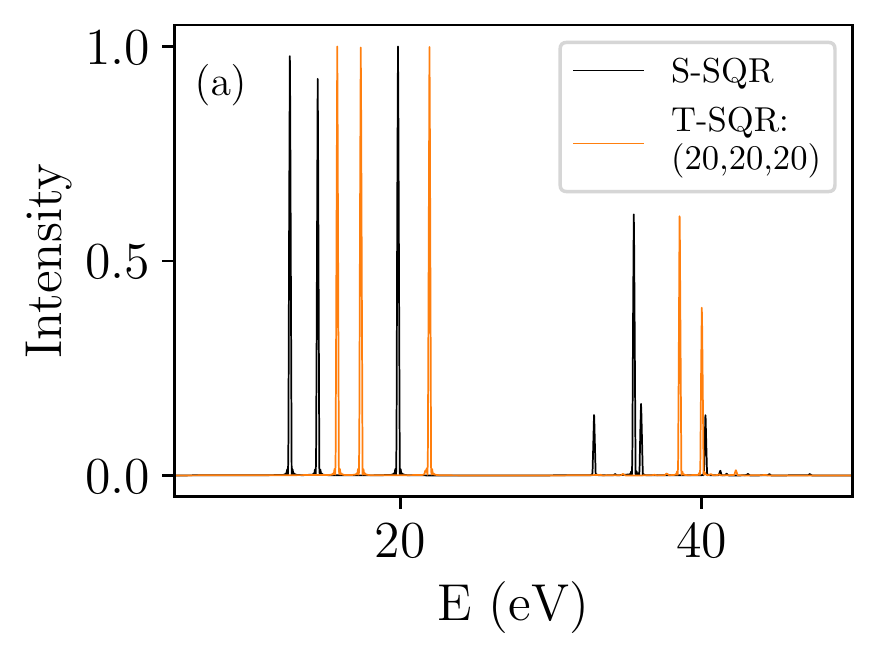}
  \includegraphics[width=5.6cm]{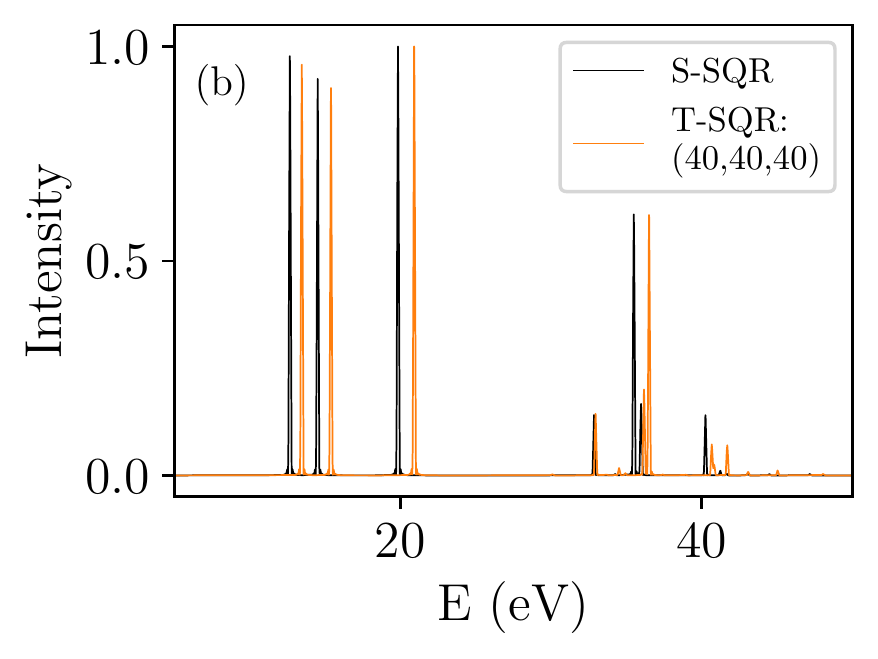}
  \includegraphics[width=5.6cm]{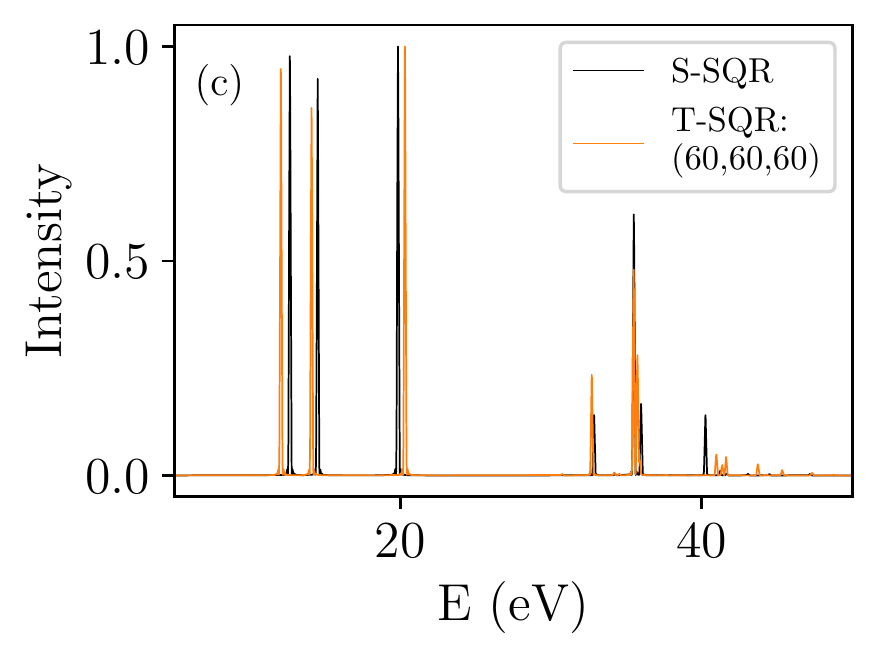}
  \includegraphics[width=5.6cm]{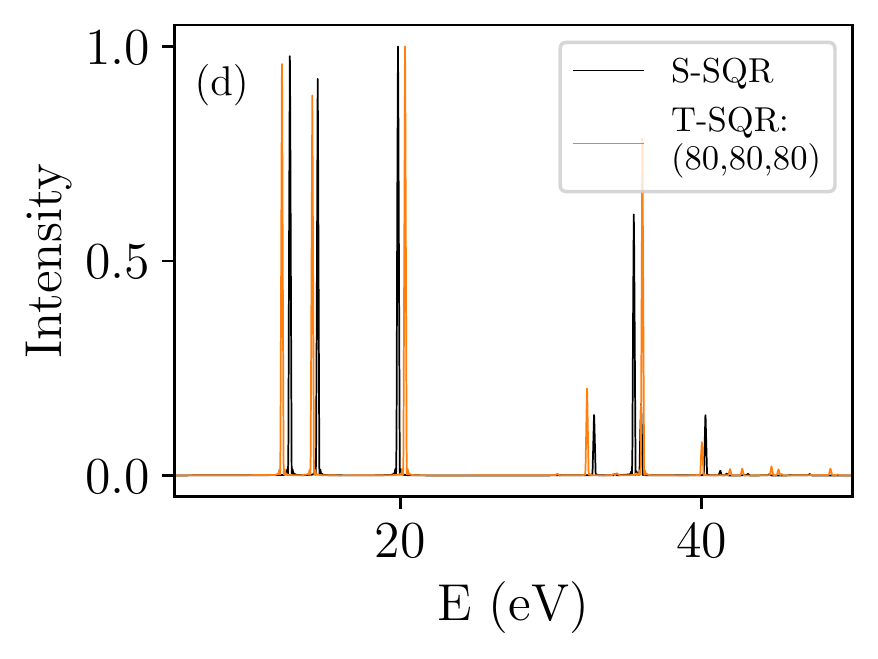}
  \includegraphics[width=5.6cm]{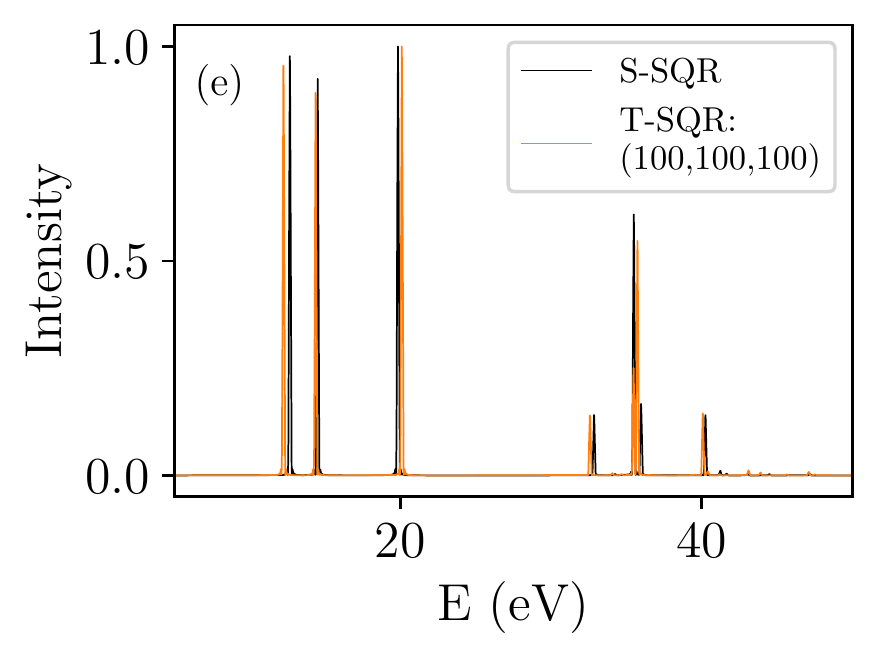}
  \includegraphics[width=5.6cm]{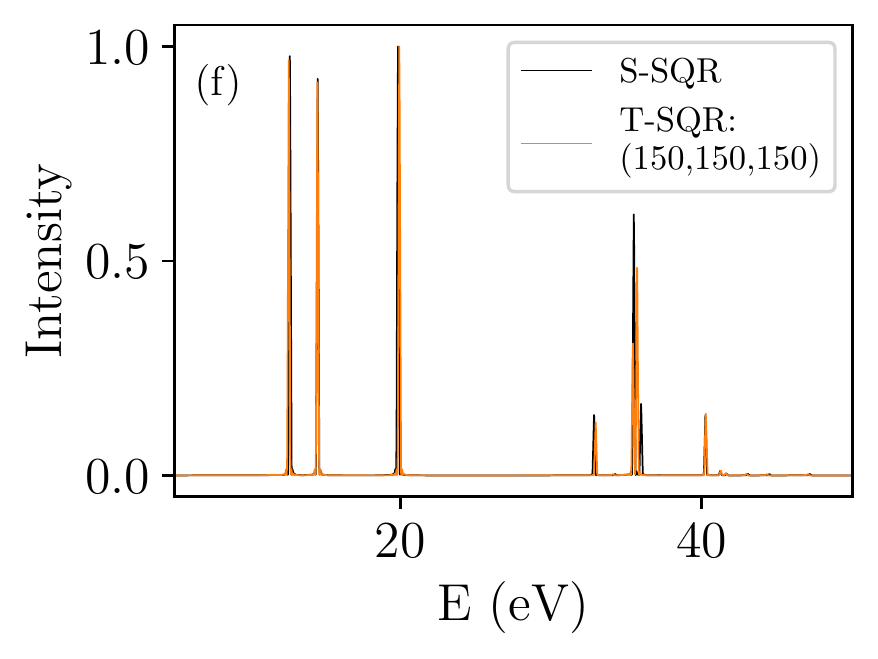}
 \caption{Singly ionized states of H$_2$O with the T-SQR Hamiltonian approach.
          Electronic eigenenergies calculated
          with Tucker rank (20,20,20), (40,40,40), (60,60,60), (80,80,80),
          (100,100,100) and (150,150,150) are shown in (a), (b), (c), (d), (e) and (f), respectively.
          Maximum rank for the T-SQR Hamiltonian is (1369, 1369, 1369).
          The spectrum in black is calculated using the S-SQR Hamiltonian approach.
          }
 \label{fig:h2o_tucker}
 \end{center}
\end{figure*}
\begin{table}[t] 
 \caption{Number of Hamiltonian terms,  memory required to store the operator
          matrices and wall-clock time for the different representation of the
          Hamiltonian. The wall-clock time is given for 20 fs time propagation.
          The calculations have been performed with 16 CPUs using shared-memory
          parallelization on the same machine and CPU type,  namely,  Dual-Core
          Intel Xeon,  processor type E5-2650 v2 running at 2.6 GHz and the
          wall-clock times are intended for their relative comparison only.
          Maximum rank for the T-SQR Hamiltonian is (1369, 1369, 1369).
          }
 \begin{ruledtabular}
  {\begin{center}
    \begin{tabular}{lcccr}
      Method & Hamil. terms & Size & Time (h:m)\\
      \hline
      SQR   & 40618 & 200 KB & 243:08 \\
      S-SQR   & 6890  & 5 MB   & 58:21 \\
      \hline
      {\bf T-SQR} \\
      (20, 20, 20)     & 800   & 7 MB   & 6:35 \\
      (40, 40, 40)     & 3200  & 11 MB  & 25:55 \\
      (60, 60, 60)     & 7200  & 26 MB  & 63:59 \\
      (80, 80, 80)     & 12800 & 82 MB  & 105.17 \\
      (100, 100, 100)  & 20000 & 90 MB  & 202:01 \\
      (150, 150, 150)  & 45000 & 132 MB & 404:29
     \end{tabular}
  \end{center} }
 \end{ruledtabular}
 \label{tab:h2o_tuck} 
\end{table}
Three FS-DOFs are formed each consisting 4 spatial (or 8 spin) orbitals.
We pruned the configurations in each FS-DOF following way:
(i) In the first FS-DOF, we allow (2$-$4) alpha, (2$-$4) beta and (6$-$8) total electrons.
(ii) In the second FS-DOF, we allow (0$-$2) alpha, (0$-$2) beta and (0$-$2) total electrons
and (iii) in the third FS-DOF, we allow (0$-$2) alpha, (0$-$2) beta and (0$-$2) total electrons.
This generates 37 configurations in each of the three FS-DOFs.
The electronic Hamiltonian is represented either as S-SQR or T-SQR Hamiltonian approach.
The initial wavefunction for the propagation is generated by applying ionization
operator on the ground electronic state of neutral H$_2$O.
The ionization operator reads
\begin{align}
 \label{eq:ionizatioOp}
 \hat{A} = \hat{a}_{_{1_\alpha}} + \hat{a}_{_{1_\beta}}
         + \hat{a}_{_{2_\alpha}} + \hat{a}_{_{2_\beta}}
         + \hat{a}_{_{3_\alpha}} + \hat{a}_{_{3_\beta}}
         + \hat{a}_{_{4_\alpha}} + \hat{a}_{_{4_\beta}} .
\end{align}
The initial wavefunction is spin doublet and overlaps with the
singly ionized states of H$_2$O.
\par
Table \ref{tab:h2o_tuck} compares the number of Hamiltonian terms, memory
required to store the operator matrices and CPU time for different representations of the electronic Hamiltonian.
We have presented two calculations with the exact Hamiltonian -
one with the original SQR Hamiltonian and another with the S-SQR
Hamiltonian where the correlated and uncorrelated operators are
summed up as much as possible.
This shows clear advantage of the S-SQR Hamiltonian over the
original SQR Hamiltonian.
\begin{figure}[t]
\begin{center}
\includegraphics[width=4.25cm]{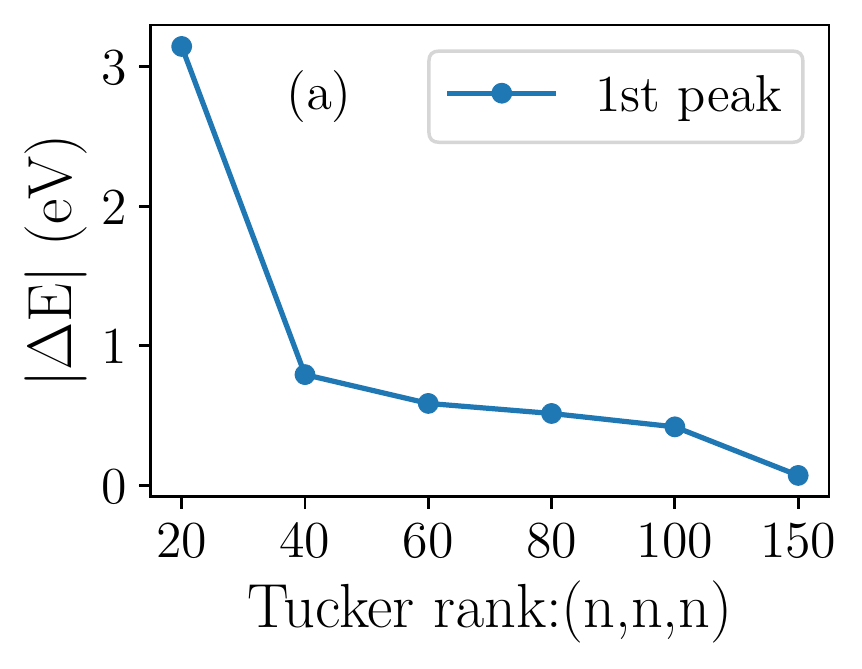}
\includegraphics[width=4.25cm]{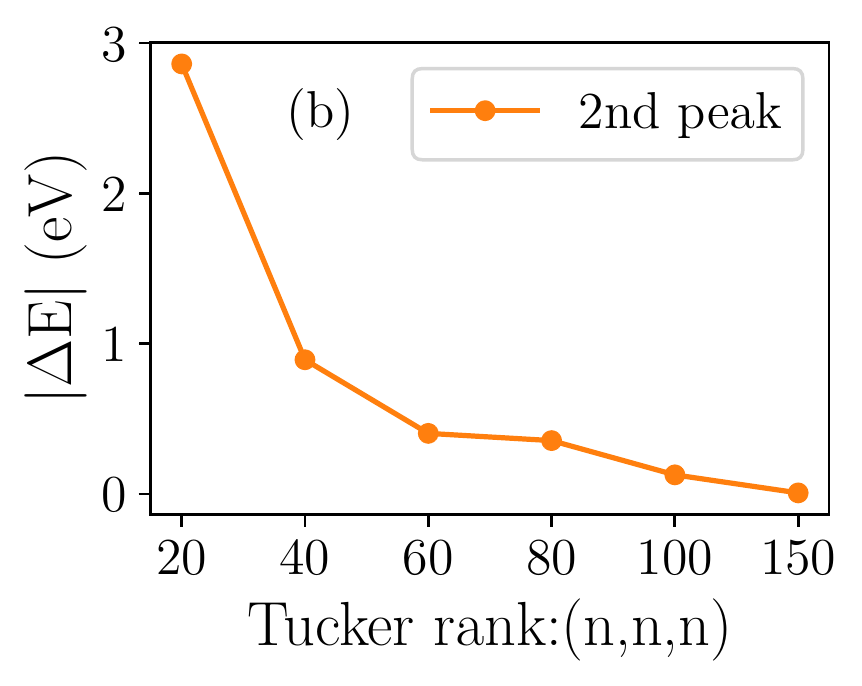}
\includegraphics[width=4.25cm]{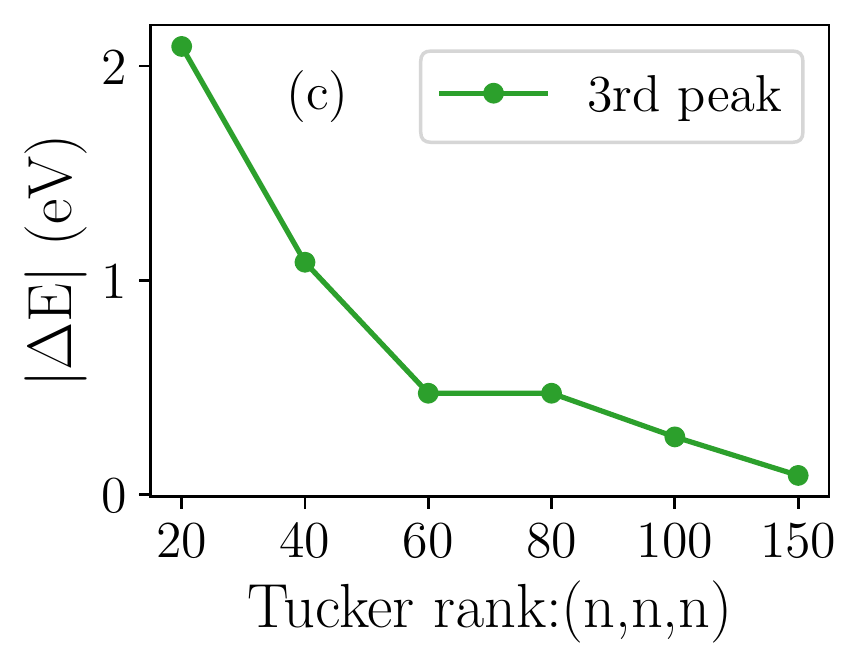}
\includegraphics[width=4.25cm]{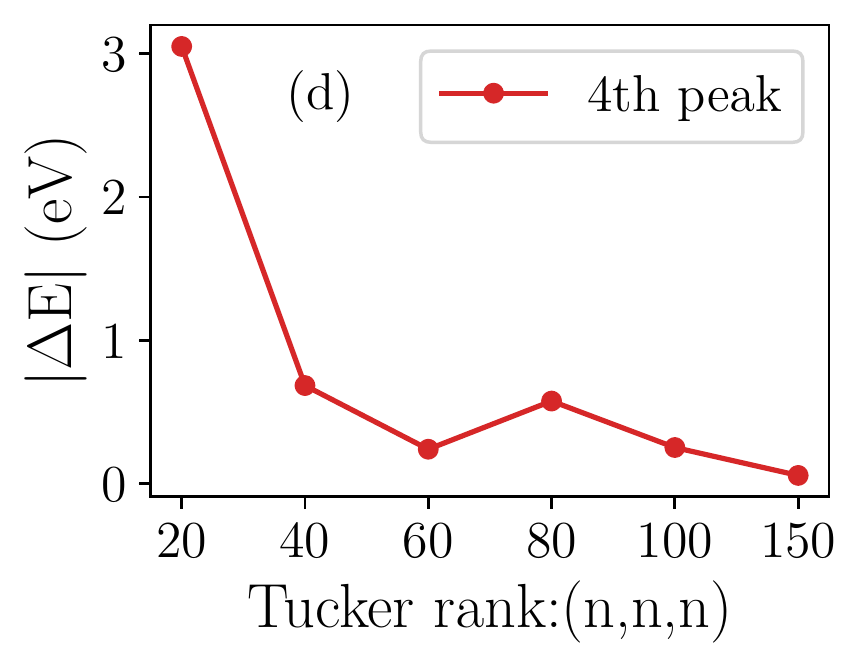}
\caption{Convergence of singly ionized state of H$_2$O with different Tucker rank.
         Difference of the electronic energies calculated with the T-SQR Hamiltonian
         from the S-SQR Hamiltonian for the 1st,
         2nd, 3rd and 4th main peak centered around 12.7 eV, 14.5 eV, 19.8 eV
         and 35.5 eV is shown in (a), (b), (c) and (d), respectively.
         }
\label{fig:h2o_tuck_conv}
\end{center}
\end{figure}
Next, we represent the electronic Hamiltonian
as T-SQR Hamiltonian with different Tucker rank.
Fig. \ref{fig:h2o_tucker} compares the electronic ionization spectrum calculated
using T-SQR Hamiltonian with S-SQR Hamiltonian.
The convergence of the four main peaks
of the singly ionized states calculated with different Tucker rank is shown in
Fig. \ref{fig:h2o_tuck_conv}.
It's clear that the the convergence is achieved with a Tucker rank much smaller
than the full rank (1369,1369,1369).
However, here the S-SQR Hamiltonian achieves a more compact form of the SQR Hamiltonian
than the converged calculation of comparable accuracy in the T-SQR approach.
\par
\begin{figure}[t]
 \begin{center}
  \includegraphics[width=8.5cm]{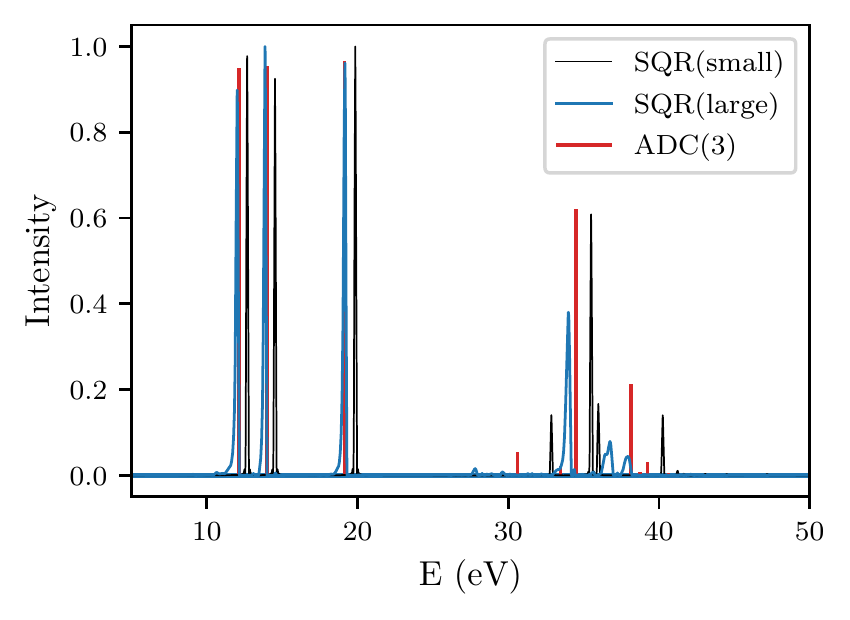}
  \caption{Comparison of electronic ionization spectrum of H$_2$O with the stick
           spectrum obtained from the ADC(3) calculation.
           SQR(small) calculation is done by allowing
           (2$-$4) alpha, (2$-$4) beta and (6$-$8) total electrons in the first FS-DOF
           and (0$-$2) alpha, (0$-$2) beta and (0$-$2) total electrons in the 2nd and
           3rd FS-DOF.
           SQR(large) calculation is done by allowing
           (0$-$4) alpha, (0$-$4) beta and (4$-$8) total electrons in the first FS-DOF
           and (0$-$4) alpha, (0$-$4) beta and (0$-$4) total electrons in the 2nd and
           3rd FS-DOF.
           }
  \label{fig:h2o_ipsprctra}
 \end{center}
\end{figure}

The ionization spectrum is compared with
the stick spectrum obtained from the
third order algebraic diagrammatic construction ADC(3) method (Fig. \ref{fig:h2o_ipsprctra}).
Two calculations in the SQR formalism are shown - one with
the FS-DOF already mentioned before (SQR(small)) and another with the FS-DOF constructed
as following (SQR(large)).
Three FS-DOFs are formed each consisting 4 spatial (or 8 spin) orbitals.
In the first FS-DOF, we allow (0$-$4) alpha, (0$-$4) beta and (4$-$8) total electrons
and in the second and third FS-DOF, we allow (0$-$4) alpha, (0$-$4) beta and (0$-$4)
total electrons. This generates 163 configurations in each of the three FS-DOFs.
The improvement in the SQR(large) from the SQR(small) calculation
is due to the larger sub-Fock space in each FS-DOF.
The first three peaks represent the valance and core-valance ionization states
of H$_2$O and correctly reproduced in the MCTDH-SQR calculation.
The ionization from the core orbital produces several main and satellites
states (spectrum above 30 eV). The small difference is due to the fact that
the configuration space used in MCTDH-SQR method is different from the ADC(3)
method.
Formally, the intensity of the peaks in the normalized SQR spectrum is in the
same limit as of the ADC(3) method. In both cases, the intensity represents the
overlap of a ionized state with the corresponding one-hole ground state.

\section{Summary and Conclusions}        \label{sec:conclusions}
In this work, we present a sum-of-products (SOP) form of the electronic
Hamiltonian in the second quantized representation (SQR). The N$^4$
scaling of the two-body part of the SQR electronic Hamiltonian
with respect to the number of spin orbitals (N) poses
a serious problem to study time dependent problems of large 
molecular systems.
We formulate two approaches to circumvent this problem,
where the primitive degrees of freedom (DOF) are represented
as Fock-space DOF (FS-DOF).
(i) First, each term of the original SQR Hamiltonian is represented
in the sub-Fock space of the corresponding FS-DOF and then the correlated
and uncorrelated terms are summed iteratively to form a more compact SOP form
of the SQR Hamiltonian (S-SQR Hamiltonian).
One obtains the most compact and still exact form of the
electronic Hamiltonian once
all terms that could be summed have been identified.
Further compactification within this strategy can only
be achieved by setting a cutoff value for the one- and two-body integrals.
This shows its intrinsic limitation as a compactification method.
(ii) Second, the electronic Hamiltonian is assumed to have a Tucker tensor form (T-SQR Hamiltonian)
and the task is to obtain the optimal core tensor and single particle operator
(SPO) matrices that minimize the difference to the
exact Hamiltonian.
Usually as the Tucker rank is much smaller than the
rank of the original Hamiltonian, one achieves a more compact SOP form of the Hamiltonian than the original tensor.

As a proof-of-concept,
we apply these two approaches to calculate the potential energy curves (PECs) of the
four lowest lying electronic states of LiH and
electronic ionization spectrum of H$_2$O.
In these two numerical examples, we compare the convergence of the electronic
eigen energies with respect to the Tucker rank. As expected, the accuracy
of the obtained results increases as the Tucker rank grows and the numerically
converged form of the
electronic Hamiltonian can be achieved with fewer product terms (lower Tucker rank)
compared to the rank of the original Hamiltonian tensor.
For LiH, the converged calculation of comparable accuracy in the T-SQR Hamiltonian approach
outperforms the S-SQR Hamiltonian approach. For the H$_2$O example, the S-SQR Hamiltonian achieves
a more compact form of the Hamiltonian than the converged calculation of
T-SQR Hamiltonian.
We believe that in problems with a larger number of FS-DOF this trend will be overturned and the numerically optimized
SOP form will be the most efficient one.

We emphasize here that our goal has been to establish whether the electronic Hamiltonian can
be systematically approximated as a SOP with an increasing accuracy as a function
of the Tucker expansion rank, and found that
this is indeed the case. We defer to future work finding practical ways to generate
SOP forms for the electronic Hamiltonian for cases where the primitive tensor operator is
too large for a direct decomposition in Tucker form.
For example, the exponential scaling with respect to the Tucker
rank can be avoided by using approximations based on the canonical
polyadic decomposition (also known as PARAFAC or CANDECOMP in literature)
of the Hamiltonian tensor. This strategy is very successful in representing
the multi-dimensional potential energy surfaces into
SOP form~\cite{Sch20:024108}, which might be extendable to the electronic problem.

%
\par
\par

\section{Supplementary Material}  \label{sec:supple}

    See the supplementary material for the tabular data of the potential
    energy curves of LiH, ionization energies of H$_2$O.

\section{Data Availability}  \label{sec:data}
    The data that support the findings of this study are available in tabular
    form in the supplementary materials.
\section{Acknowledgments}

    We thank Prof. H.-D. Meyer for his
    important assistance with the MCTDH calculations.
    The authors thank JUSTUS 2 in Ulm for computing time.
    The authors declare no conflicts of interest.

\appendix
\section{Compacted form of the electronic Hamiltonian in the spin orbital basis}
 \label{ap:ham_spinorb}
The electronic Hamiltonian written in Eq. \ref{eq:ham_el_jw} reads as
\begin{align}
 \hat{H}_{el} = \hat{H}_{1} + \hat{H}_{2},
\end{align}
where
\begin{align}
 \label{eq:ham_el1_jw}
 \hat{H}_{1} = &
    \sum_{ij} h_{_{ij}}
        \left(
        \prod_{q=a+1}^{b-1} \hat{\sigma}^z_q
        \right)
        \hat{\sigma}_i^{+} \hat{\sigma}_j^{-} \nonumber\\
             = &
        \sum_{ij} h_{ij} \hat{O}_{ij} ,
\end{align}
and
\begin{align}
 \label{eq:ham_el2_jw}
 \hat{H}_{2} = &
 \frac{1}{2}\sum_{ijkl} v_{_{ijkl}}
          \left( \prod_{q=a+1}^{b-1} \hat{\sigma}^z_q
              \prod_{q^\prime=c+1}^{d-1} \hat{\sigma}^z_{q^\prime}
          \right) \nonumber\\
       &   \mathrm{sgn}(j-i)\mathrm{sgn}(l-k)
    \hat{\sigma}_i^{+} \hat{\sigma}_j^{+} \hat{\sigma}_l^{-} \hat{\sigma}_k^{-} \nonumber\\
    = &
    \sum_{ijkl} v_{ijkl} \hat{O}_{ijkl} .
\end{align}
Here, $\hat{O}_{ij}$ and $\hat{O}_{ijkl}$ denote the entire operator form of
the 1e- and 2e-part of Eqs.~\ref{eq:ham_el1_jw} and \ref{eq:ham_el1_jw}, respectively.
Now, one can divide the Hamiltonian by the index of the orbital belongs to
their corresponding FS-DOF.
\begin{align}
 \hat{H}_{1} = \sum_{\alpha}^{f} \sum_{i_\alpha j_\alpha}^{m_\alpha}
               h_{i_\alpha j_\alpha} \hat{O}_{i_\alpha j_\alpha}
             + \sum_{\alpha < \beta}^{f} \mathbf{P}(\alpha, \beta)
               \sum_{i_\alpha}^{m_\alpha} \sum_{j_\beta}^{m_\beta}
               h_{i_\alpha j_\beta} \hat{O}_{i_\alpha j_\beta} ,
\end{align}
where $i_\kappa$ denotes the index of the orbital of the $\kappa$-th FS-DOF
and $m_\kappa$ is the number of spin orbitals grouped to form the $\kappa$-th
FS-DOF and $f$ is the number of FS-DOFs under consideration.
$\mathbf{P}(\alpha, \beta)$ is the permutation operator that generates
all the possible permutation of the indices
\begin{align}
 \mathbf{P}(\alpha, \beta) X(\alpha, \beta) = X(\alpha, \beta) + X(\beta, \alpha) .
\end{align}
Now, one can divide the operator $\hat{O}_{i_\alpha j_\beta}$ into sub-operators
that act only on the corresponding FS-DOF.
The one-body part of the Hamiltonian reads
\begin{widetext}
\begin{align}
 \hat{H}_{1} &= \sum_{\alpha}^{f} 
                \sum_{i_\alpha j_\alpha}^{m_\alpha} h_{i_\alpha j_\alpha}
                \hat{O}^{(\alpha)}_{i_\alpha j_\alpha} 
             + \sum_{\alpha < \beta}^{f} \mathbf{P}(\alpha, \beta)
                \sum_{i_\alpha}^{m_\alpha} \sum_{j_\beta}^{m_\beta}
                h_{i_\alpha j_\beta} \hat{O}^{(\alpha)}_{i_\alpha}
                \hat{O}^{(\alpha+1)} \cdots \hat{O}^{(\beta-1)}
                \hat{O}^{(\beta)}_{j_\beta} \nonumber\\
             &= \sum_{\alpha}^{f} 
                \left(
                       \sum_{i_\alpha j_\alpha}^{m_\alpha}
                       h_{i_\alpha j_\alpha} \hat{O}^{(\alpha)}_{i_\alpha j_\alpha}
                \right) 
             + \sum_{\alpha < \beta}^{f} \mathbf{P}(\alpha, \beta)
                \sum_{i_\alpha}^{m_\alpha}
                \hat{O}^{(\alpha)}_{i_\alpha}
                \hat{O}^{(\alpha+1)} \cdots \hat{O}^{(\beta-1)}
                \left(
                \sum_{j_\beta}^{m_\beta}
                h_{i_\alpha j_\beta} 
                \hat{O}^{(\beta)}_{j_\beta}
                \right) .
\end{align}
\end{widetext}
Here, $\hat{O}^{(\kappa)}_{a_\kappa b_\kappa \cdots}$ denotes an operator string consists of
spin ladder operators of ($a_\kappa, b_\kappa, \cdots$) orbitals and sign change or identity
operator of other orbitals of the $\kappa$-th FS-DOF. $\hat{O}^{\kappa}$ denotes an operator
string consists of either sign change or identity operator of the spin orbitals of $\kappa$-th FS-DOF.
Similarly, the two-body part of the Hamiltonian reads
\begin{widetext}
\begin{align}
 \hat{H}_{2} &= \sum_{ijkl} v_{ijkl} \hat{O}_{ijkl} \nonumber\\
             &= \sum_{\alpha}^{f}
                \sum_{i_\alpha j_\alpha k_\alpha l_\alpha}^{m_\alpha}
                v_{i_\alpha j_\alpha k_\alpha l_\alpha}
                \hat{O}^{(\alpha)}_{i_\alpha j_\alpha k_\alpha l_\alpha} \nonumber\\
             &+ \sum_{\alpha < \beta}^{f} \mathbf{P}(\alpha, \alpha, \alpha, \beta)
                \sum_{i_\alpha j_\alpha k_\alpha}^{m_\alpha} \sum_{l_\beta}^{m_\beta}
                v_{i_\alpha j_\alpha k_\alpha l_\beta}
                \hat{O}^{(\alpha)}_{i_\alpha j_\alpha k_\alpha}
                \hat{O}^{(\alpha+1)} \cdots \hat{O}^{(\beta-1)}
                \hat{O}^{(\beta)}_{l_\beta} \nonumber\\
             &+ \sum_{\alpha < \beta}^{f} \mathbf{P}(\alpha, \alpha, \beta, \beta)
                \sum_{i_\alpha j_\alpha}^{m_\alpha} \sum_{k_\beta l_\beta}^{m_\beta}
                v_{i_\alpha j_\alpha k_\beta l_\beta}
                \hat{O}^{(\alpha)}_{i_\alpha j_\alpha}
                \hat{O}^{(\beta)}_{k_\beta l_\beta} \nonumber\\
             &+ \sum_{\alpha < \beta < \gamma}^{f} \mathbf{P}(\alpha, \alpha, \beta, \gamma)
                \sum_{i_\alpha j_\alpha}^{m_\alpha} \sum_{k_\beta}^{m_\beta}
                \sum_{l_\gamma}^{m_\gamma}
                v_{i_\alpha j_\alpha k_\beta l_\gamma}
                \hat{O}^{(\alpha)}_{i_\alpha j_\alpha} \hat{O}^{(\beta)}_{k_\beta}
                \hat{O}^{(\beta+1)} \cdots \hat{O}^{(\gamma-1)}
                \hat{O}^{(\gamma)}_{l_\gamma} \nonumber\\
             &+ \sum_{\alpha < \beta < \gamma < \delta}^{f} \mathbf{P}(\alpha, \beta, \gamma, \delta)
                \sum_{i_\alpha}^{m_\alpha} \sum_{j_\beta}^{m_\beta}
                \sum_{k_\gamma}^{m_\gamma} \sum_{l_\delta}^{m_\delta}
                v_{i_\alpha j_\beta k_\gamma l_\delta}
                \hat{O}^{(\alpha)}_{i_\alpha}
                \hat{O}^{(\alpha+1)} \cdots \hat{O}^{(\beta-1)}
                \hat{O}^{(\beta)}_{j_\beta} \hat{O}^{(\gamma)}_{k_\gamma}
                \hat{O}^{(\gamma+1)} \cdots \hat{O}^{(\delta-1)}
                \hat{O}^{(\delta)}_{l_\delta} \nonumber\\
             &= \sum_{\alpha}^{f}
                \left(
                      \sum_{i_\alpha j_\alpha k_\alpha l_\alpha}^{m_\alpha}
                      v_{i_\alpha j_\alpha k_\alpha l_\alpha}
                      \hat{O}^{(\alpha)}_{i_\alpha j_\alpha k_\alpha l_\alpha}
                \right) \nonumber\\
             &+ \sum_{\alpha < \beta}^{f} \mathbf{P}(\alpha, \alpha, \alpha, \beta)
                \left(
                      \sum_{i_\alpha j_\alpha k_\alpha}^{m_\alpha}
                      v_{i_\alpha j_\alpha k_\alpha l_\beta}
                      \hat{O}^{(\alpha)}_{i_\alpha j_\alpha k_\alpha}
                \right)
                \hat{O}^{(\alpha+1)} \cdots \hat{O}^{(\beta-1)}
                \sum_{l_\beta}^{m_\beta}
                \hat{O}^{(\beta)}_{l_\beta} \nonumber\\
             &+ \sum_{\alpha < \beta}^{f} \mathbf{P}(\alpha, \alpha, \beta, \beta)
                \sum_{i_\alpha j_\alpha}^{m_\alpha}
                \hat{O}^{(\alpha)}_{i_\alpha j_\alpha}
                \left(
                      \sum_{k_\beta l_\beta}^{m_\beta}
                      v_{i_\alpha j_\alpha k_\beta l_\beta}
                      \hat{O}^{(\beta)}_{k_\beta l_\beta}
                      \right) \nonumber\\
             &+ \sum_{\alpha < \beta < \gamma}^{f} \mathbf{P}(\alpha, \alpha, \beta, \gamma)
                \left(
                      \sum_{i_\alpha j_\alpha}^{m_\alpha}
                      v_{i_\alpha j_\alpha k_\beta l_\gamma}
                      \hat{O}^{(\alpha)}_{i_\alpha j_\alpha}
                \right)
                \sum_{k_\beta}^{m_\beta}
                \sum_{l_\gamma}^{m_\gamma}
                \hat{O}^{(\beta)}_{k_\beta}
                \hat{O}^{(\beta+1)} \cdots \hat{O}^{(\gamma-1)}
                \hat{O}^{(\gamma)}_{l_\gamma} \nonumber\\
             &+ \sum_{\alpha < \beta < \gamma < \delta}^{f} \mathbf{P}(\alpha, \beta, \gamma, \delta)
                \sum_{i_\alpha}^{m_\alpha} \sum_{j_\beta}^{m_\beta}
                \sum_{k_\gamma}^{m_\gamma}
                \hat{O}^{(\alpha)}_{i_\alpha}
                \hat{O}^{(\alpha+1)} \cdots \hat{O}^{(\beta-1)}
                \hat{O}^{(\beta)}_{j_\beta} \hat{O}^{(\gamma)}_{k_\gamma}
                \hat{O}^{(\gamma+1)} \cdots \hat{O}^{(\delta-1)}
                \left(
                      \sum_{l_\delta}^{m_\delta}
                      v_{i_\alpha j_\beta k_\gamma l_\delta}
                      \hat{O}^{(\delta)}_{l_\delta}
                \right) .
\end{align}
\end{widetext}
Here, we can sum the terms in the parenthesis as they act only on one primitive
DOF to reduce the number of Hamiltonian terms.
\section{Application of Hamiltonian on the wavefunction: Matrix vs Mapping}
\label{ap:matvecmul}
The operators acting on the SPFs of the FS-DOF are very sparse as each term
the electronic Hamiltonian given in Eq.~\ref{eq:ham_el_jw} can connect
one bra configuration of the FS-DOF to only one specific ket configuration
of the corresponding FS-DOF. Thus, the matrix operators have the
property that at most there is one non-zero entry in each row and column that
can be either equal to 1 or -1. As a consequence, for an $(m \times m)$ matrix, at most
$m$ entries are different from $0$. This sparsity can be exploited
by matrix-vector multiplication algorithms with linear rather that quadratic
scaling.
For example, $a_1^{\dagger}a_4$ operator term is applied to a sequence
of four combined S-DOFs as
\begin{align}
 a_1^{\dagger}a_4 = \sigma_1^{\dagger} S_2 S_3 \sigma_4 = 
 \begin{pmatrix}
    0 & 0 \\
    1 & 0 \\
 \end{pmatrix}_{(1)}
 \begin{pmatrix}
    1 & 0 \\
    0 & -1 \\
 \end{pmatrix}_{(2)}
 \begin{pmatrix}
    1 & 0 \\
    0 & -1 \\
 \end{pmatrix}_{(3)}
 \begin{pmatrix}
    0 & 1 \\
    0 & 0 \\
 \end{pmatrix}_{(4)} .
\end{align}
The same operator in the FS-DOF sub-Fock space reads
\begin{align}
 a_1^{\dagger}a_4 = \sigma_1^{\dagger} S_2 S_3 \sigma_4 = 
 {\tiny
 \begin{pmatrix}
    0 & 0 & 0 & 0 & 0 & 0 & 0 & 0 & 0 & 0 & 0 & 0 & 0 & 0 & 0 & 0\\
    0 & 0 & 0 & 0 & 0 & 0 & 0 & 0 & 0 & 0 & 0 & 0 & 0 & 0 & 0 & 0\\
    0 & 0 & 0 & 1 & 0 & 0 & 0 & 0 & 0 & 0 & 0 & 0 & 0 & 0 & 0 & 0\\
    0 & 0 & 0 & 0 & 0 & 0 & 0 & 0 & 0 & 0 & 0 & 0 & 0 & 0 & 0 & 0\\
    0 & 0 & 0 & 0 & 0 & 0 & 0 & 0 & 0 & 0 & 0 & 0 & 0 & 0 & 0 & 0\\
    0 & 0 & 0 & 0 & 0 & 0 & -1 & 0 & 0 & 0 & 0 & 0 & 0 & 0 & 0 & 0\\
    0 & 0 & 0 & 0 & 0 & 0 & 0 & 0 & 0 & 0 & 0 & 0 & 0 & 0 & 0 & 0\\
    0 & 0 & 0 & 0 & 0 & 0 & 0 & 0 & 0 & 0 & 0 & 0 & 0 & 0 & 0 & 0\\
    0 & 0 & 0 & 0 & 0 & 0 & 0 & 0 & 0 & 0 & 0 & 0 & 0 & 0 & 0 & 0\\
    0 & 0 & 0 & 0 & 0 & 0 & 0 & 0 & 0 & 0 & -1 & 0 & 0 & 0 & 0 & 0\\
    0 & 0 & 0 & 0 & 0 & 0 & 0 & 0 & 0 & 0 & 0 & 0 & 0 & 0 & 0 & 0\\
    0 & 0 & 0 & 0 & 0 & 0 & 0 & 0 & 0 & 0 & 0 & 0 & 0 & 0 & 0 & 0\\
    0 & 0 & 0 & 0 & 0 & 0 & 0 & 0 & 0 & 0 & 0 & 0 & 0 & 1 & 0 & 0\\
    0 & 0 & 0 & 0 & 0 & 0 & 0 & 0 & 0 & 0 & 0 & 0 & 0 & 0 & 0 & 0\\
    0 & 0 & 0 & 0 & 0 & 0 & 0 & 0 & 0 & 0 & 0 & 0 & 0 & 0 & 0 & 0\\
    0 & 0 & 0 & 0 & 0 & 0 & 0 & 0 & 0 & 0 & 0 & 0 & 0 & 0 & 0 & 0
 \end{pmatrix}.
 }
\end{align}
So, the application of this operator to the SPFs of the FS-DOF operator
is a mapping where only certain memory positions need to be copied
\begin{align}
 {\tiny
 \begin{pmatrix}
    0 & 0 & 0 & 0 & 0 & 0 & 0 & 0 & 0 & 0 & 0 & 0 & 0 & 0 & 0 & 0\\
    0 & 0 & 0 & 0 & 0 & 0 & 0 & 0 & 0 & 0 & 0 & 0 & 0 & 0 & 0 & 0\\
    0 & 0 & 0 & 1 & 0 & 0 & 0 & 0 & 0 & 0 & 0 & 0 & 0 & 0 & 0 & 0\\
    0 & 0 & 0 & 0 & 0 & 0 & 0 & 0 & 0 & 0 & 0 & 0 & 0 & 0 & 0 & 0\\
    0 & 0 & 0 & 0 & 0 & 0 & 0 & 0 & 0 & 0 & 0 & 0 & 0 & 0 & 0 & 0\\
    0 & 0 & 0 & 0 & 0 & 0 & -1 & 0 & 0 & 0 & 0 & 0 & 0 & 0 & 0 & 0\\
    0 & 0 & 0 & 0 & 0 & 0 & 0 & 0 & 0 & 0 & 0 & 0 & 0 & 0 & 0 & 0\\
    0 & 0 & 0 & 0 & 0 & 0 & 0 & 0 & 0 & 0 & 0 & 0 & 0 & 0 & 0 & 0\\
    0 & 0 & 0 & 0 & 0 & 0 & 0 & 0 & 0 & 0 & 0 & 0 & 0 & 0 & 0 & 0\\
    0 & 0 & 0 & 0 & 0 & 0 & 0 & 0 & 0 & 0 & -1 & 0 & 0 & 0 & 0 & 0\\
    0 & 0 & 0 & 0 & 0 & 0 & 0 & 0 & 0 & 0 & 0 & 0 & 0 & 0 & 0 & 0\\
    0 & 0 & 0 & 0 & 0 & 0 & 0 & 0 & 0 & 0 & 0 & 0 & 0 & 0 & 0 & 0\\
    0 & 0 & 0 & 0 & 0 & 0 & 0 & 0 & 0 & 0 & 0 & 0 & 0 & 1 & 0 & 0\\
    0 & 0 & 0 & 0 & 0 & 0 & 0 & 0 & 0 & 0 & 0 & 0 & 0 & 0 & 0 & 0\\
    0 & 0 & 0 & 0 & 0 & 0 & 0 & 0 & 0 & 0 & 0 & 0 & 0 & 0 & 0 & 0\\
    0 & 0 & 0 & 0 & 0 & 0 & 0 & 0 & 0 & 0 & 0 & 0 & 0 & 0 & 0 & 0
 \end{pmatrix}
 \begin{pmatrix}
    a_{1} \\
    a_{2} \\
    a_{3} \\
    a_{4} \\
    a_{5} \\
    a_{6} \\
    a_{7} \\
    a_{8} \\
    a_{9} \\
    a_{10} \\
    a_{11} \\
    a_{12} \\
    a_{13} \\
    a_{14} \\
    a_{15} \\
    a_{16}
 \end{pmatrix}
 =
 \begin{pmatrix}
    0 \\
    0 \\
    a_{4} \\
    0 \\
    0 \\
    -a_{7} \\
    0 \\
    0 \\
    0 \\
    -a_{11} \\
    0 \\
    0 \\
    a_{14} \\
    0 \\
    0 \\
    0
 \end{pmatrix}.
 }
\end{align}
In this case, instead of $16^2$ floating point operations (FLOP), the mapping
strategy requires just $4$ FLOPs.
Another advantage of the mapping algorithm is that one requires significantly
less memory for the matrix-vector multiplication $-$ i.e., only the indices of
rows, columns and the sign of the non-zero matrix elements need to be stored
instead of the full matrix.
The pseudo-code for the mapping algorithm is given in Algorithm~\ref{al:mapping}.
\begin{algorithm}
 \label{al:mapping}
 \caption{Mapping algorithm}
 \SetKwInOut{KwIn}{Input}
 \SetKwInOut{KwOut}{Output}
 \KwIn{Arrays: ${\bf row}(i), {\bf col}(i), {\bf sgn}(i), {\bf ovec}(i)$,
               $i=1, 2, \cdots, n$, where each element of
               ${\bf row}$ and ${\bf col}$ is integer,
               ${\bf ovec}$ is real/complex
               and ${\bf sgn}$ is either $1$ or $-1$.
               ${\bf row}, {\bf col}, {\bf sgn}$ stores the indices of
               row and column and sign of the non-zero matrix elements
               of the operator, respectively.
               ${\bf ovec}$ stores the coefficients of vector before multiplication.
       }
 \KwOut{Array: ${\bf nvec}(i)$, $i=1, 2, \cdots, m$, where each element of
               ${\bf nvec}$ is a real/complex.
               ${\bf nvec}$ stores the resulting vector of the matrix-vector multiplication.
       }
 \tcc{Initialization of ${\bf nvec}$ as zero}
 ${\bf nvec}$ = 0.0 \\
 \tcc{map the elements of {\bf nvec} to the elements of {\bf ovec} with proper sign}
 \For{$i \leftarrow 0$ \KwTo $n$}{
    \hspace{1em} ${\bf nvec}({\bf row}(i)) = {\bf sgn}(i)*{\bf ovec}({\bf col}(i))$
 }
\end{algorithm}
This has been implemented in the Heidelberg MCTDH package and can be switched on
whenever very sparse operators are encountered for any degree of freedom, not only
in the MCTDH-SQR context.


\begin{thebibliography}{61}%
\makeatletter
\providecommand \@ifxundefined [1]{%
 \@ifx{#1\undefined}
}%
\providecommand \@ifnum [1]{%
 \ifnum #1\expandafter \@firstoftwo
 \else \expandafter \@secondoftwo
 \fi
}%
\providecommand \@ifx [1]{%
 \ifx #1\expandafter \@firstoftwo
 \else \expandafter \@secondoftwo
 \fi
}%
\providecommand \natexlab [1]{#1}%
\providecommand \enquote  [1]{``#1''}%
\providecommand \bibnamefont  [1]{#1}%
\providecommand \bibfnamefont [1]{#1}%
\providecommand \citenamefont [1]{#1}%
\providecommand \href@noop [0]{\@secondoftwo}%
\providecommand \href [0]{\begingroup \@sanitize@url \@href}%
\providecommand \@href[1]{\@@startlink{#1}\@@href}%
\providecommand \@@href[1]{\endgroup#1\@@endlink}%
\providecommand \@sanitize@url [0]{\catcode `\\12\catcode `\$12\catcode
  `\&12\catcode `\#12\catcode `\^12\catcode `\_12\catcode `\%12\relax}%
\providecommand \@@startlink[1]{}%
\providecommand \@@endlink[0]{}%
\providecommand \url  [0]{\begingroup\@sanitize@url \@url }%
\providecommand \@url [1]{\endgroup\@href {#1}{\urlprefix }}%
\providecommand \urlprefix  [0]{URL }%
\providecommand \Eprint [0]{\href }%
\providecommand \doibase [0]{http://dx.doi.org/}%
\providecommand \selectlanguage [0]{\@gobble}%
\providecommand \bibinfo  [0]{\@secondoftwo}%
\providecommand \bibfield  [0]{\@secondoftwo}%
\providecommand \translation [1]{[#1]}%
\providecommand \BibitemOpen [0]{}%
\providecommand \bibitemStop [0]{}%
\providecommand \bibitemNoStop [0]{.\EOS\space}%
\providecommand \EOS [0]{\spacefactor3000\relax}%
\providecommand \BibitemShut  [1]{\csname bibitem#1\endcsname}%
\let\auto@bib@innerbib\@empty
\bibitem [{\citenamefont {Meyer}, \citenamefont {Manthe},\ and\ \citenamefont
  {Cederbaum}(1990)}]{mey90:73}%
  \BibitemOpen
  \bibfield  {author} {\bibinfo {author} {\bibfnamefont {H.~D.}\ \bibnamefont
  {Meyer}}, \bibinfo {author} {\bibfnamefont {U.}~\bibnamefont {Manthe}}, \
  and\ \bibinfo {author} {\bibfnamefont {L.~S.}\ \bibnamefont {Cederbaum}},\
  }\href {\doibase 10.1016/0009-2614(90)87014-i} {\bibfield  {journal}
  {\bibinfo  {journal} {Chem. Phys. Lett.}\ }\textbf {\bibinfo {volume}
  {165}},\ \bibinfo {pages} {73} (\bibinfo {year} {1990})}\BibitemShut
  {NoStop}%
\bibitem [{\citenamefont {Manthe}, \citenamefont {Meyer},\ and\ \citenamefont
  {Cederbaum}(1992)}]{man92:3199}%
  \BibitemOpen
  \bibfield  {author} {\bibinfo {author} {\bibfnamefont {U.}~\bibnamefont
  {Manthe}}, \bibinfo {author} {\bibfnamefont {H.-D.}\ \bibnamefont {Meyer}}, \
  and\ \bibinfo {author} {\bibfnamefont {L.~S.}\ \bibnamefont {Cederbaum}},\
  }\href {\doibase 10.1063/1.463007} {\bibfield  {journal} {\bibinfo  {journal}
  {J. Chem. Phys.}\ }\textbf {\bibinfo {volume} {97}},\ \bibinfo {pages} {3199}
  (\bibinfo {year} {1992})}\BibitemShut {NoStop}%
\bibitem [{\citenamefont {Beck}\ \emph {et~al.}(2000)\citenamefont {Beck},
  \citenamefont {J\"ackle}, \citenamefont {Worth},\ and\ \citenamefont
  {Meyer}}]{bec00:1}%
  \BibitemOpen
  \bibfield  {author} {\bibinfo {author} {\bibfnamefont {M.~H.}\ \bibnamefont
  {Beck}}, \bibinfo {author} {\bibfnamefont {A.}~\bibnamefont {J\"ackle}},
  \bibinfo {author} {\bibfnamefont {G.~A.}\ \bibnamefont {Worth}}, \ and\
  \bibinfo {author} {\bibfnamefont {H.-D.}\ \bibnamefont {Meyer}},\ }\href
  {\doibase 10.1016/S0370-1573(99)00047-2} {\bibfield  {journal} {\bibinfo
  {journal} {Phys. Rep.}\ }\textbf {\bibinfo {volume} {324}},\ \bibinfo {pages}
  {1} (\bibinfo {year} {2000})}\BibitemShut {NoStop}%
\bibitem [{\citenamefont {Meyer}\ and\ \citenamefont
  {Worth}(2003)}]{mey03:251}%
  \BibitemOpen
  \bibfield  {author} {\bibinfo {author} {\bibfnamefont {H.-D.}\ \bibnamefont
  {Meyer}}\ and\ \bibinfo {author} {\bibfnamefont {G.~A.}\ \bibnamefont
  {Worth}},\ }\href {\doibase 10.1007/s00214-003-0439-1} {\bibfield  {journal}
  {\bibinfo  {journal} {Theor. Chem. Acc.}\ }\textbf {\bibinfo {volume}
  {109}},\ \bibinfo {pages} {251} (\bibinfo {year} {2003})}\BibitemShut
  {NoStop}%
\bibitem [{\citenamefont {Meyer}, \citenamefont {Gatti},\ and\ \citenamefont
  {Worth}(2009)}]{mey09:book}%
  \BibitemOpen
  \bibinfo {editor} {\bibfnamefont {H.-D.}\ \bibnamefont {Meyer}}, \bibinfo
  {editor} {\bibfnamefont {F.}~\bibnamefont {Gatti}}, \ and\ \bibinfo {editor}
  {\bibfnamefont {G.~A.}\ \bibnamefont {Worth}},\ eds.,\ \href {\doibase
  10.1002/9783527627400} {\emph {\bibinfo {title} {Multidimensional Quantum
  Dynamics}}}\ (\bibinfo  {publisher} {Wiley},\ \bibinfo {year}
  {2009})\BibitemShut {NoStop}%
\bibitem [{\citenamefont {Wang}\ and\ \citenamefont
  {Thoss}(2003)}]{wan03:1289}%
  \BibitemOpen
  \bibfield  {author} {\bibinfo {author} {\bibfnamefont {H.}~\bibnamefont
  {Wang}}\ and\ \bibinfo {author} {\bibfnamefont {M.}~\bibnamefont {Thoss}},\
  }\href {\doibase 10.1063/1.1580111} {\bibfield  {journal} {\bibinfo
  {journal} {J. Chem. Phys.}\ }\textbf {\bibinfo {volume} {119}},\ \bibinfo
  {pages} {1289} (\bibinfo {year} {2003})}\BibitemShut {NoStop}%
\bibitem [{\citenamefont {Manthe}(2008)}]{man08:164116}%
  \BibitemOpen
  \bibfield  {author} {\bibinfo {author} {\bibfnamefont {U.}~\bibnamefont
  {Manthe}},\ }\href {\doibase 10.1063/1.2902982} {\bibfield  {journal}
  {\bibinfo  {journal} {J. Chem. Phys.}\ }\textbf {\bibinfo {volume} {128}},\
  \bibinfo {pages} {164116} (\bibinfo {year} {2008})}\BibitemShut {NoStop}%
\bibitem [{\citenamefont {Vendrell}\ and\ \citenamefont
  {Meyer}(2011)}]{Ven11:44135}%
  \BibitemOpen
  \bibfield  {author} {\bibinfo {author} {\bibfnamefont {O.}~\bibnamefont
  {Vendrell}}\ and\ \bibinfo {author} {\bibfnamefont {H.-D.}\ \bibnamefont
  {Meyer}},\ }\href {\doibase 10.1063/1.3535541} {\bibfield  {journal}
  {\bibinfo  {journal} {J. Chem. Phys.}\ }\textbf {\bibinfo {volume} {134}},\
  \bibinfo {pages} {044135} (\bibinfo {year} {2011})}\BibitemShut {NoStop}%
\bibitem [{\citenamefont {Meyer}(2012)}]{mey12:351}%
  \BibitemOpen
  \bibfield  {author} {\bibinfo {author} {\bibfnamefont {H.-D.}\ \bibnamefont
  {Meyer}},\ }\href {\doibase 10.1002/wcms.87} {\bibfield  {journal} {\bibinfo
  {journal} {WIREs Comput Mol Sci}\ }\textbf {\bibinfo {volume} {2}},\ \bibinfo
  {pages} {351} (\bibinfo {year} {2012})}\BibitemShut {NoStop}%
\bibitem [{\citenamefont {Wang}(2015)}]{wan15:7951}%
  \BibitemOpen
  \bibfield  {author} {\bibinfo {author} {\bibfnamefont {H.}~\bibnamefont
  {Wang}},\ }\href {\doibase 10.1021/acs.jpca.5b03256} {\bibfield  {journal}
  {\bibinfo  {journal} {J. Phys. Chem. A}\ }\textbf {\bibinfo {volume} {119}},\
  \bibinfo {pages} {7951} (\bibinfo {year} {2015})}\BibitemShut {NoStop}%
\bibitem [{\citenamefont {Caillat}\ \emph {et~al.}(2005)\citenamefont
  {Caillat}, \citenamefont {Zanghellini}, \citenamefont {Kitzler},
  \citenamefont {Koch}, \citenamefont {Kreuzer},\ and\ \citenamefont
  {Scrinzi}}]{cai05:12712}%
  \BibitemOpen
  \bibfield  {author} {\bibinfo {author} {\bibfnamefont {J.}~\bibnamefont
  {Caillat}}, \bibinfo {author} {\bibfnamefont {J.}~\bibnamefont
  {Zanghellini}}, \bibinfo {author} {\bibfnamefont {M.}~\bibnamefont
  {Kitzler}}, \bibinfo {author} {\bibfnamefont {O.}~\bibnamefont {Koch}},
  \bibinfo {author} {\bibfnamefont {W.}~\bibnamefont {Kreuzer}}, \ and\
  \bibinfo {author} {\bibfnamefont {A.}~\bibnamefont {Scrinzi}},\ }\href
  {\doibase 10.1103/physreva.71.012712} {\bibfield  {journal} {\bibinfo
  {journal} {Phys. Rev. A}\ }\textbf {\bibinfo {volume} {71}},\ \bibinfo
  {pages} {012712} (\bibinfo {year} {2005})}\BibitemShut {NoStop}%
\bibitem [{\citenamefont {Alon}, \citenamefont {Streltsov},\ and\ \citenamefont
  {Cederbaum}(2007)}]{alo07:154103}%
  \BibitemOpen
  \bibfield  {author} {\bibinfo {author} {\bibfnamefont {O.~E.}\ \bibnamefont
  {Alon}}, \bibinfo {author} {\bibfnamefont {A.~I.}\ \bibnamefont {Streltsov}},
  \ and\ \bibinfo {author} {\bibfnamefont {L.~S.}\ \bibnamefont {Cederbaum}},\
  }\href {\doibase 10.1063/1.2771159} {\bibfield  {journal} {\bibinfo
  {journal} {J. Chem. Phys.}\ }\textbf {\bibinfo {volume} {127}},\ \bibinfo
  {pages} {154103} (\bibinfo {year} {2007})}\BibitemShut {NoStop}%
\bibitem [{\citenamefont {Hochstuhl}\ and\ \citenamefont
  {Bonitz}(2011)}]{hoc11:084106}%
  \BibitemOpen
  \bibfield  {author} {\bibinfo {author} {\bibfnamefont {D.}~\bibnamefont
  {Hochstuhl}}\ and\ \bibinfo {author} {\bibfnamefont {M.}~\bibnamefont
  {Bonitz}},\ }\href {\doibase 10.1063/1.3553176} {\bibfield  {journal}
  {\bibinfo  {journal} {J. Chem. Phys.}\ }\textbf {\bibinfo {volume} {134}},\
  \bibinfo {pages} {084106} (\bibinfo {year} {2011})}\BibitemShut {NoStop}%
\bibitem [{\citenamefont {Sato}\ and\ \citenamefont
  {Ishikawa}(2013)}]{sat13:023402}%
  \BibitemOpen
  \bibfield  {author} {\bibinfo {author} {\bibfnamefont {T.}~\bibnamefont
  {Sato}}\ and\ \bibinfo {author} {\bibfnamefont {K.~L.}\ \bibnamefont
  {Ishikawa}},\ }\href {\doibase 10.1103/physreva.88.023402} {\bibfield
  {journal} {\bibinfo  {journal} {Phys. Rev. A}\ }\textbf {\bibinfo {volume}
  {88}},\ \bibinfo {pages} {023402} (\bibinfo {year} {2013})}\BibitemShut
  {NoStop}%
\bibitem [{\citenamefont {Lode}\ \emph {et~al.}(2020)\citenamefont {Lode},
  \citenamefont {L{\'{e}}v{\^{e}}que}, \citenamefont {Madsen}, \citenamefont
  {Streltsov},\ and\ \citenamefont {Alon}}]{lod20:011001}%
  \BibitemOpen
  \bibfield  {author} {\bibinfo {author} {\bibfnamefont {A.~U.}\ \bibnamefont
  {Lode}}, \bibinfo {author} {\bibfnamefont {C.}~\bibnamefont
  {L{\'{e}}v{\^{e}}que}}, \bibinfo {author} {\bibfnamefont {L.~B.}\
  \bibnamefont {Madsen}}, \bibinfo {author} {\bibfnamefont {A.~I.}\
  \bibnamefont {Streltsov}}, \ and\ \bibinfo {author} {\bibfnamefont {O.~E.}\
  \bibnamefont {Alon}},\ }\href {\doibase 10.1103/revmodphys.92.011001}
  {\bibfield  {journal} {\bibinfo  {journal} {Rev. Mod. Phys.}\ }\textbf
  {\bibinfo {volume} {92}},\ \bibinfo {pages} {011001} (\bibinfo {year}
  {2020})}\BibitemShut {NoStop}%
\bibitem [{\citenamefont {Alon}, \citenamefont {Streltsov},\ and\ \citenamefont
  {Cederbaum}(2008)}]{alo08:33613}%
  \BibitemOpen
  \bibfield  {author} {\bibinfo {author} {\bibfnamefont {O.~E.}\ \bibnamefont
  {Alon}}, \bibinfo {author} {\bibfnamefont {A.~I.}\ \bibnamefont {Streltsov}},
  \ and\ \bibinfo {author} {\bibfnamefont {L.~S.}\ \bibnamefont {Cederbaum}},\
  }\href {\doibase 10.1103/physreva.77.033613} {\bibfield  {journal} {\bibinfo
  {journal} {Phys. Rev. A}\ }\textbf {\bibinfo {volume} {77}},\ \bibinfo
  {pages} {033613} (\bibinfo {year} {2008})}\BibitemShut {NoStop}%
\bibitem [{\citenamefont {Kr{\"{o}}nke}\ \emph {et~al.}(2013)\citenamefont
  {Kr{\"{o}}nke}, \citenamefont {Cao}, \citenamefont {Vendrell},\ and\
  \citenamefont {Schmelcher}}]{kroe13:63018}%
  \BibitemOpen
  \bibfield  {author} {\bibinfo {author} {\bibfnamefont {S.}~\bibnamefont
  {Kr{\"{o}}nke}}, \bibinfo {author} {\bibfnamefont {L.}~\bibnamefont {Cao}},
  \bibinfo {author} {\bibfnamefont {O.}~\bibnamefont {Vendrell}}, \ and\
  \bibinfo {author} {\bibfnamefont {P.}~\bibnamefont {Schmelcher}},\ }\href
  {\doibase 10.1088/1367-2630/15/6/063018} {\bibfield  {journal} {\bibinfo
  {journal} {New J. Phys.}\ }\textbf {\bibinfo {volume} {15}},\ \bibinfo
  {pages} {063018} (\bibinfo {year} {2013})}\BibitemShut {NoStop}%
\bibitem [{\citenamefont {Wang}\ and\ \citenamefont
  {Thoss}(2009)}]{wan09:024114}%
  \BibitemOpen
  \bibfield  {author} {\bibinfo {author} {\bibfnamefont {H.}~\bibnamefont
  {Wang}}\ and\ \bibinfo {author} {\bibfnamefont {M.}~\bibnamefont {Thoss}},\
  }\href {\doibase 10.1063/1.3173823} {\bibfield  {journal} {\bibinfo
  {journal} {J. Chem. Phys.}\ }\textbf {\bibinfo {volume} {131}},\ \bibinfo
  {pages} {024114} (\bibinfo {year} {2009})}\BibitemShut {NoStop}%
\bibitem [{\citenamefont {Balzer}\ \emph {et~al.}(2015)\citenamefont {Balzer},
  \citenamefont {Li}, \citenamefont {Vendrell},\ and\ \citenamefont
  {Eckstein}}]{Bal15:45136}%
  \BibitemOpen
  \bibfield  {author} {\bibinfo {author} {\bibfnamefont {K.}~\bibnamefont
  {Balzer}}, \bibinfo {author} {\bibfnamefont {Z.}~\bibnamefont {Li}}, \bibinfo
  {author} {\bibfnamefont {O.}~\bibnamefont {Vendrell}}, \ and\ \bibinfo
  {author} {\bibfnamefont {M.}~\bibnamefont {Eckstein}},\ }\href {\doibase
  10.1103/PhysRevB.91.045136} {\bibfield  {journal} {\bibinfo  {journal} {Phys.
  Rev. B}\ }\textbf {\bibinfo {volume} {91}},\ \bibinfo {pages} {045136}
  (\bibinfo {year} {2015})}\BibitemShut {NoStop}%
\bibitem [{\citenamefont {Wang}\ \emph {et~al.}(2011)\citenamefont {Wang},
  \citenamefont {Pshenichnyuk}, \citenamefont {H\"{a}rtle},\ and\ \citenamefont
  {Thoss}}]{Wan11:244506}%
  \BibitemOpen
  \bibfield  {author} {\bibinfo {author} {\bibfnamefont {H.}~\bibnamefont
  {Wang}}, \bibinfo {author} {\bibfnamefont {I.}~\bibnamefont {Pshenichnyuk}},
  \bibinfo {author} {\bibfnamefont {R.}~\bibnamefont {H\"{a}rtle}}, \ and\
  \bibinfo {author} {\bibfnamefont {M.}~\bibnamefont {Thoss}},\ }\href
  {\doibase 10.1063/1.3660206} {\bibfield  {journal} {\bibinfo  {journal} {J.
  Chem. Phys.}\ }\textbf {\bibinfo {volume} {135}},\ \bibinfo {pages} {244506}
  (\bibinfo {year} {2011})}\BibitemShut {NoStop}%
\bibitem [{\citenamefont {Wang}\ and\ \citenamefont
  {Thoss}(2013{\natexlab{a}})}]{wan13:134704}%
  \BibitemOpen
  \bibfield  {author} {\bibinfo {author} {\bibfnamefont {H.}~\bibnamefont
  {Wang}}\ and\ \bibinfo {author} {\bibfnamefont {M.}~\bibnamefont {Thoss}},\
  }\href {\doibase 10.1063/1.4798404} {\bibfield  {journal} {\bibinfo
  {journal} {J. Chem. Phys.}\ }\textbf {\bibinfo {volume} {138}},\ \bibinfo
  {pages} {134704} (\bibinfo {year} {2013}{\natexlab{a}})}\BibitemShut
  {NoStop}%
\bibitem [{\citenamefont {Wang}\ and\ \citenamefont
  {Thoss}(2013{\natexlab{b}})}]{wan13:7431}%
  \BibitemOpen
  \bibfield  {author} {\bibinfo {author} {\bibfnamefont {H.}~\bibnamefont
  {Wang}}\ and\ \bibinfo {author} {\bibfnamefont {M.}~\bibnamefont {Thoss}},\
  }\href {\doibase 10.1021/jp401464b} {\bibfield  {journal} {\bibinfo
  {journal} {J. Phys. Chem. A}\ }\textbf {\bibinfo {volume} {117}},\ \bibinfo
  {pages} {7431} (\bibinfo {year} {2013}{\natexlab{b}})}\BibitemShut {NoStop}%
\bibitem [{\citenamefont {Wilner}\ \emph {et~al.}(2013)\citenamefont {Wilner},
  \citenamefont {Wang}, \citenamefont {Cohen}, \citenamefont {Thoss},\ and\
  \citenamefont {Rabani}}]{Wil13:045137}%
  \BibitemOpen
  \bibfield  {author} {\bibinfo {author} {\bibfnamefont {E.~Y.}\ \bibnamefont
  {Wilner}}, \bibinfo {author} {\bibfnamefont {H.}~\bibnamefont {Wang}},
  \bibinfo {author} {\bibfnamefont {G.}~\bibnamefont {Cohen}}, \bibinfo
  {author} {\bibfnamefont {M.}~\bibnamefont {Thoss}}, \ and\ \bibinfo {author}
  {\bibfnamefont {E.}~\bibnamefont {Rabani}},\ }\href {\doibase
  10.1103/physrevb.88.045137} {\bibfield  {journal} {\bibinfo  {journal} {Phys.
  Rev. B}\ }\textbf {\bibinfo {volume} {88}},\ \bibinfo {pages} {045137}
  (\bibinfo {year} {2013})}\BibitemShut {NoStop}%
\bibitem [{\citenamefont {Wilner}\ \emph {et~al.}(2014)\citenamefont {Wilner},
  \citenamefont {Wang}, \citenamefont {Thoss},\ and\ \citenamefont
  {Rabani}}]{Wil14:205129}%
  \BibitemOpen
  \bibfield  {author} {\bibinfo {author} {\bibfnamefont {E.~Y.}\ \bibnamefont
  {Wilner}}, \bibinfo {author} {\bibfnamefont {H.}~\bibnamefont {Wang}},
  \bibinfo {author} {\bibfnamefont {M.}~\bibnamefont {Thoss}}, \ and\ \bibinfo
  {author} {\bibfnamefont {E.}~\bibnamefont {Rabani}},\ }\href {\doibase
  10.1103/physrevb.89.205129} {\bibfield  {journal} {\bibinfo  {journal} {Phys.
  Rev. B}\ }\textbf {\bibinfo {volume} {89}},\ \bibinfo {pages} {205129}
  (\bibinfo {year} {2014})}\BibitemShut {NoStop}%
\bibitem [{\citenamefont {Manthe}\ and\ \citenamefont
  {Weike}(2017)}]{man17:064117}%
  \BibitemOpen
  \bibfield  {author} {\bibinfo {author} {\bibfnamefont {U.}~\bibnamefont
  {Manthe}}\ and\ \bibinfo {author} {\bibfnamefont {T.}~\bibnamefont {Weike}},\
  }\href {\doibase 10.1063/1.4975662} {\bibfield  {journal} {\bibinfo
  {journal} {J. Chem. Phys.}\ }\textbf {\bibinfo {volume} {146}},\ \bibinfo
  {pages} {064117} (\bibinfo {year} {2017})}\BibitemShut {NoStop}%
\bibitem [{\citenamefont {Weike}\ and\ \citenamefont
  {Manthe}(2020)}]{wei20:034101}%
  \BibitemOpen
  \bibfield  {author} {\bibinfo {author} {\bibfnamefont {T.}~\bibnamefont
  {Weike}}\ and\ \bibinfo {author} {\bibfnamefont {U.}~\bibnamefont {Manthe}},\
  }\href {\doibase 10.1063/1.5140984} {\bibfield  {journal} {\bibinfo
  {journal} {J. Chem. Phys.}\ }\textbf {\bibinfo {volume} {152}},\ \bibinfo
  {pages} {034101} (\bibinfo {year} {2020})}\BibitemShut {NoStop}%
\bibitem [{\citenamefont {Sasmal}\ and\ \citenamefont
  {Vendrell}(2020)}]{Sas20:154110}%
  \BibitemOpen
  \bibfield  {author} {\bibinfo {author} {\bibfnamefont {S.}~\bibnamefont
  {Sasmal}}\ and\ \bibinfo {author} {\bibfnamefont {O.}~\bibnamefont
  {Vendrell}},\ }\href {\doibase 10.1063/5.0028116} {\bibfield  {journal}
  {\bibinfo  {journal} {J. Chem. Phys.}\ }\textbf {\bibinfo {volume} {153}},\
  \bibinfo {pages} {154110} (\bibinfo {year} {2020})}\BibitemShut {NoStop}%
\bibitem [{\citenamefont {White}(1992)}]{Whi92:2863}%
  \BibitemOpen
  \bibfield  {author} {\bibinfo {author} {\bibfnamefont {S.~R.}\ \bibnamefont
  {White}},\ }\href {\doibase 10.1103/physrevlett.69.2863} {\bibfield
  {journal} {\bibinfo  {journal} {Physical Review Letters}\ }\textbf {\bibinfo
  {volume} {69}},\ \bibinfo {pages} {2863} (\bibinfo {year}
  {1992})}\BibitemShut {NoStop}%
\bibitem [{\citenamefont {White}(1993)}]{Whi93:10345}%
  \BibitemOpen
  \bibfield  {author} {\bibinfo {author} {\bibfnamefont {S.~R.}\ \bibnamefont
  {White}},\ }\href {\doibase 10.1103/physrevb.48.10345} {\bibfield  {journal}
  {\bibinfo  {journal} {Physical Review B}\ }\textbf {\bibinfo {volume} {48}},\
  \bibinfo {pages} {10345} (\bibinfo {year} {1993})}\BibitemShut {NoStop}%
\bibitem [{\citenamefont {Schollwoeck}(2005)}]{sch05:259}%
  \BibitemOpen
  \bibfield  {author} {\bibinfo {author} {\bibfnamefont {U.}~\bibnamefont
  {Schollwoeck}},\ }\href {\doibase 10.1007/bfb0106064} {\bibfield  {journal}
  {\bibinfo  {journal} {Rev. Mod. Phys.}\ }\textbf {\bibinfo {volume} {77}},\
  \bibinfo {pages} {259} (\bibinfo {year} {2005})}\BibitemShut {NoStop}%
\bibitem [{\citenamefont {Schollw\"{o}ck}(2011)}]{Sch11:96}%
  \BibitemOpen
  \bibfield  {author} {\bibinfo {author} {\bibfnamefont {U.}~\bibnamefont
  {Schollw\"{o}ck}},\ }\href {\doibase 10.1016/j.aop.2010.09.012} {\bibfield
  {journal} {\bibinfo  {journal} {Annals of Physics}\ }\textbf {\bibinfo
  {volume} {326}},\ \bibinfo {pages} {96} (\bibinfo {year} {2011})}\BibitemShut
  {NoStop}%
\bibitem [{\citenamefont {Chan}\ and\ \citenamefont
  {Sharma}(2011)}]{Cha11:465}%
  \BibitemOpen
  \bibfield  {author} {\bibinfo {author} {\bibfnamefont {G.~K.-L.}\
  \bibnamefont {Chan}}\ and\ \bibinfo {author} {\bibfnamefont {S.}~\bibnamefont
  {Sharma}},\ }\href {\doibase 10.1146/annurev-physchem-032210-103338}
  {\bibfield  {journal} {\bibinfo  {journal} {Annual Review of Physical
  Chemistry}\ }\textbf {\bibinfo {volume} {62}},\ \bibinfo {pages} {465}
  (\bibinfo {year} {2011})}\BibitemShut {NoStop}%
\bibitem [{\citenamefont {McCulloch}(2007)}]{McC07:P10014}%
  \BibitemOpen
  \bibfield  {author} {\bibinfo {author} {\bibfnamefont {I.~P.}\ \bibnamefont
  {McCulloch}},\ }\href {\doibase 10.1088/1742-5468/2007/10/p10014} {\bibfield
  {journal} {\bibinfo  {journal} {Journal of Statistical Mechanics: Theory and
  Experiment}\ }\textbf {\bibinfo {volume} {2007}},\ \bibinfo {pages} {P10014}
  (\bibinfo {year} {2007})}\BibitemShut {NoStop}%
\bibitem [{\citenamefont {Chan}\ \emph {et~al.}(2016)\citenamefont {Chan},
  \citenamefont {Keselman}, \citenamefont {Nakatani}, \citenamefont {Li},\ and\
  \citenamefont {White}}]{Cha16:014102}%
  \BibitemOpen
  \bibfield  {author} {\bibinfo {author} {\bibfnamefont {G.~K.-L.}\
  \bibnamefont {Chan}}, \bibinfo {author} {\bibfnamefont {A.}~\bibnamefont
  {Keselman}}, \bibinfo {author} {\bibfnamefont {N.}~\bibnamefont {Nakatani}},
  \bibinfo {author} {\bibfnamefont {Z.}~\bibnamefont {Li}}, \ and\ \bibinfo
  {author} {\bibfnamefont {S.~R.}\ \bibnamefont {White}},\ }\href {\doibase
  10.1063/1.4955108} {\bibfield  {journal} {\bibinfo  {journal} {J. Chem.
  Phys.}\ }\textbf {\bibinfo {volume} {145}},\ \bibinfo {pages} {014102}
  (\bibinfo {year} {2016})}\BibitemShut {NoStop}%
\bibitem [{\citenamefont {Keller}\ \emph {et~al.}(2015)\citenamefont {Keller},
  \citenamefont {Dolfi}, \citenamefont {Troyer},\ and\ \citenamefont
  {Reiher}}]{Kel15:244118}%
  \BibitemOpen
  \bibfield  {author} {\bibinfo {author} {\bibfnamefont {S.}~\bibnamefont
  {Keller}}, \bibinfo {author} {\bibfnamefont {M.}~\bibnamefont {Dolfi}},
  \bibinfo {author} {\bibfnamefont {M.}~\bibnamefont {Troyer}}, \ and\ \bibinfo
  {author} {\bibfnamefont {M.}~\bibnamefont {Reiher}},\ }\href {\doibase
  10.1063/1.4939000} {\bibfield  {journal} {\bibinfo  {journal} {J. Chem.
  Phys.}\ }\textbf {\bibinfo {volume} {143}},\ \bibinfo {pages} {244118}
  (\bibinfo {year} {2015})}\BibitemShut {NoStop}%
\bibitem [{\citenamefont {Yanai}\ \emph {et~al.}(2014)\citenamefont {Yanai},
  \citenamefont {Kurashige}, \citenamefont {Mizukami}, \citenamefont
  {Chalupsk{\'{y}}}, \citenamefont {Lan},\ and\ \citenamefont
  {Saitow}}]{Yan14:283}%
  \BibitemOpen
  \bibfield  {author} {\bibinfo {author} {\bibfnamefont {T.}~\bibnamefont
  {Yanai}}, \bibinfo {author} {\bibfnamefont {Y.}~\bibnamefont {Kurashige}},
  \bibinfo {author} {\bibfnamefont {W.}~\bibnamefont {Mizukami}}, \bibinfo
  {author} {\bibfnamefont {J.}~\bibnamefont {Chalupsk{\'{y}}}}, \bibinfo
  {author} {\bibfnamefont {T.~N.}\ \bibnamefont {Lan}}, \ and\ \bibinfo
  {author} {\bibfnamefont {M.}~\bibnamefont {Saitow}},\ }\href {\doibase
  10.1002/qua.24808} {\bibfield  {journal} {\bibinfo  {journal} {International
  Journal of Quantum Chemistry}\ }\textbf {\bibinfo {volume} {115}},\ \bibinfo
  {pages} {283} (\bibinfo {year} {2014})}\BibitemShut {NoStop}%
\bibitem [{\citenamefont {J\"{a}ckle}\ and\ \citenamefont
  {Meyer}(1996)}]{jaec96:7974}%
  \BibitemOpen
  \bibfield  {author} {\bibinfo {author} {\bibfnamefont {A.}~\bibnamefont
  {J\"{a}ckle}}\ and\ \bibinfo {author} {\bibfnamefont {H.-D.}\ \bibnamefont
  {Meyer}},\ }\href {\doibase 10.1063/1.471513} {\bibfield  {journal} {\bibinfo
   {journal} {J. Chem. Phys.}\ }\textbf {\bibinfo {volume} {104}},\ \bibinfo
  {pages} {7974} (\bibinfo {year} {1996})}\BibitemShut {NoStop}%
\bibitem [{\citenamefont {J\"{a}ckle}\ and\ \citenamefont
  {Meyer}(1998)}]{jaec98:3772}%
  \BibitemOpen
  \bibfield  {author} {\bibinfo {author} {\bibfnamefont {A.}~\bibnamefont
  {J\"{a}ckle}}\ and\ \bibinfo {author} {\bibfnamefont {H.-D.}\ \bibnamefont
  {Meyer}},\ }\href {\doibase 10.1063/1.476977} {\bibfield  {journal} {\bibinfo
   {journal} {J. Chem. Phys.}\ }\textbf {\bibinfo {volume} {109}},\ \bibinfo
  {pages} {3772} (\bibinfo {year} {1998})}\BibitemShut {NoStop}%
\bibitem [{\citenamefont {Pel{\'{a}}ez}\ and\ \citenamefont
  {Meyer}(2013)}]{Pel13:014108}%
  \BibitemOpen
  \bibfield  {author} {\bibinfo {author} {\bibfnamefont {D.}~\bibnamefont
  {Pel{\'{a}}ez}}\ and\ \bibinfo {author} {\bibfnamefont {H.-D.}\ \bibnamefont
  {Meyer}},\ }\href {\doibase 10.1063/1.4773021} {\bibfield  {journal}
  {\bibinfo  {journal} {J. Chem. Phys.}\ }\textbf {\bibinfo {volume} {138}},\
  \bibinfo {pages} {014108} (\bibinfo {year} {2013})}\BibitemShut {NoStop}%
\bibitem [{\citenamefont {Schr\"{o}der}\ and\ \citenamefont
  {Meyer}(2017)}]{Sch17:064105}%
  \BibitemOpen
  \bibfield  {author} {\bibinfo {author} {\bibfnamefont {M.}~\bibnamefont
  {Schr\"{o}der}}\ and\ \bibinfo {author} {\bibfnamefont {H.-D.}\ \bibnamefont
  {Meyer}},\ }\href {\doibase 10.1063/1.4991851} {\bibfield  {journal}
  {\bibinfo  {journal} {J. Chem. Phys.}\ }\textbf {\bibinfo {volume} {147}},\
  \bibinfo {pages} {064105} (\bibinfo {year} {2017})}\BibitemShut {NoStop}%
\bibitem [{\citenamefont {Otto}, \citenamefont {Chiang},\ and\ \citenamefont
  {Pel{\'{a}}ez}(2018)}]{Ott18:116}%
  \BibitemOpen
  \bibfield  {author} {\bibinfo {author} {\bibfnamefont {F.}~\bibnamefont
  {Otto}}, \bibinfo {author} {\bibfnamefont {Y.-C.}\ \bibnamefont {Chiang}}, \
  and\ \bibinfo {author} {\bibfnamefont {D.}~\bibnamefont {Pel{\'{a}}ez}},\
  }\href {\doibase 10.1016/j.chemphys.2017.11.013} {\bibfield  {journal}
  {\bibinfo  {journal} {Chemical Physics}\ }\textbf {\bibinfo {volume} {509}},\
  \bibinfo {pages} {116} (\bibinfo {year} {2018})}\BibitemShut {NoStop}%
\bibitem [{\citenamefont {Otto}(2014)}]{ott14:14106}%
  \BibitemOpen
  \bibfield  {author} {\bibinfo {author} {\bibfnamefont {F.}~\bibnamefont
  {Otto}},\ }\href {\doibase 10.1063/1.4856135} {\bibfield  {journal} {\bibinfo
   {journal} {J. Chem. Phys.}\ }\textbf {\bibinfo {volume} {140}},\ \bibinfo
  {pages} {014106} (\bibinfo {year} {2014})}\BibitemShut {NoStop}%
\bibitem [{\citenamefont {Schr\"{o}der}(2020)}]{Sch20:024108}%
  \BibitemOpen
  \bibfield  {author} {\bibinfo {author} {\bibfnamefont {M.}~\bibnamefont
  {Schr\"{o}der}},\ }\href {\doibase 10.1063/1.5140085} {\bibfield  {journal}
  {\bibinfo  {journal} {J. Chem. Phys.}\ }\textbf {\bibinfo {volume} {152}},\
  \bibinfo {pages} {024108} (\bibinfo {year} {2020})}\BibitemShut {NoStop}%
\bibitem [{\citenamefont {Manzhos}\ \emph {et~al.}(2005)\citenamefont
  {Manzhos}, \citenamefont {Wang}, \citenamefont {Dawes},\ and\ \citenamefont
  {Carrington}}]{Man05:5295}%
  \BibitemOpen
  \bibfield  {author} {\bibinfo {author} {\bibfnamefont {S.}~\bibnamefont
  {Manzhos}}, \bibinfo {author} {\bibfnamefont {X.}~\bibnamefont {Wang}},
  \bibinfo {author} {\bibfnamefont {R.}~\bibnamefont {Dawes}}, \ and\ \bibinfo
  {author} {\bibfnamefont {T.}~\bibnamefont {Carrington}},\ }\href {\doibase
  10.1021/jp055253z} {\bibfield  {journal} {\bibinfo  {journal} {J. Phys. Chem.
  A}\ }\textbf {\bibinfo {volume} {110}},\ \bibinfo {pages} {5295} (\bibinfo
  {year} {2005})}\BibitemShut {NoStop}%
\bibitem [{\citenamefont {Manzhos}\ and\ \citenamefont
  {Carrington}(2006{\natexlab{a}})}]{Man06:084109}%
  \BibitemOpen
  \bibfield  {author} {\bibinfo {author} {\bibfnamefont {S.}~\bibnamefont
  {Manzhos}}\ and\ \bibinfo {author} {\bibfnamefont {T.}~\bibnamefont
  {Carrington}},\ }\href {\doibase 10.1063/1.2336223} {\bibfield  {journal}
  {\bibinfo  {journal} {J. Chem. Phys.}\ }\textbf {\bibinfo {volume} {125}},\
  \bibinfo {pages} {084109} (\bibinfo {year} {2006}{\natexlab{a}})}\BibitemShut
  {NoStop}%
\bibitem [{\citenamefont {Manzhos}\ and\ \citenamefont
  {Carrington}(2006{\natexlab{b}})}]{Man06:194105}%
  \BibitemOpen
  \bibfield  {author} {\bibinfo {author} {\bibfnamefont {S.}~\bibnamefont
  {Manzhos}}\ and\ \bibinfo {author} {\bibfnamefont {T.}~\bibnamefont
  {Carrington}},\ }\href {\doibase 10.1063/1.2387950} {\bibfield  {journal}
  {\bibinfo  {journal} {J. Chem. Phys.}\ }\textbf {\bibinfo {volume} {125}},\
  \bibinfo {pages} {194105} (\bibinfo {year} {2006}{\natexlab{b}})}\BibitemShut
  {NoStop}%
\bibitem [{\citenamefont {Koch}\ and\ \citenamefont
  {Zhang}(2014)}]{Koc14:021101}%
  \BibitemOpen
  \bibfield  {author} {\bibinfo {author} {\bibfnamefont {W.}~\bibnamefont
  {Koch}}\ and\ \bibinfo {author} {\bibfnamefont {D.~H.}\ \bibnamefont
  {Zhang}},\ }\href {\doibase 10.1063/1.4887508} {\bibfield  {journal}
  {\bibinfo  {journal} {J. Chem. Phys.}\ }\textbf {\bibinfo {volume} {141}},\
  \bibinfo {pages} {021101} (\bibinfo {year} {2014})}\BibitemShut {NoStop}%
\bibitem [{\citenamefont {Shen}\ \emph {et~al.}(2015)\citenamefont {Shen},
  \citenamefont {Chen}, \citenamefont {Zhang}, \citenamefont {Shao},\ and\
  \citenamefont {Zhang}}]{She15:144701}%
  \BibitemOpen
  \bibfield  {author} {\bibinfo {author} {\bibfnamefont {X.}~\bibnamefont
  {Shen}}, \bibinfo {author} {\bibfnamefont {J.}~\bibnamefont {Chen}}, \bibinfo
  {author} {\bibfnamefont {Z.}~\bibnamefont {Zhang}}, \bibinfo {author}
  {\bibfnamefont {K.}~\bibnamefont {Shao}}, \ and\ \bibinfo {author}
  {\bibfnamefont {D.~H.}\ \bibnamefont {Zhang}},\ }\href {\doibase
  10.1063/1.4932226} {\bibfield  {journal} {\bibinfo  {journal} {J. Chem.
  Phys.}\ }\textbf {\bibinfo {volume} {143}},\ \bibinfo {pages} {144701}
  (\bibinfo {year} {2015})}\BibitemShut {NoStop}%
\bibitem [{\citenamefont {Pradhan}\ and\ \citenamefont
  {Brown}(2016{\natexlab{a}})}]{Pra16:174305}%
  \BibitemOpen
  \bibfield  {author} {\bibinfo {author} {\bibfnamefont {E.}~\bibnamefont
  {Pradhan}}\ and\ \bibinfo {author} {\bibfnamefont {A.}~\bibnamefont
  {Brown}},\ }\href {\doibase 10.1063/1.4948440} {\bibfield  {journal}
  {\bibinfo  {journal} {J. Chem. Phys.}\ }\textbf {\bibinfo {volume} {144}},\
  \bibinfo {pages} {174305} (\bibinfo {year} {2016}{\natexlab{a}})}\BibitemShut
  {NoStop}%
\bibitem [{\citenamefont {Pradhan}\ and\ \citenamefont
  {Brown}(2016{\natexlab{b}})}]{Pra16:158}%
  \BibitemOpen
  \bibfield  {author} {\bibinfo {author} {\bibfnamefont {E.}~\bibnamefont
  {Pradhan}}\ and\ \bibinfo {author} {\bibfnamefont {A.}~\bibnamefont
  {Brown}},\ }\href {\doibase 10.1016/j.jms.2016.06.009} {\bibfield  {journal}
  {\bibinfo  {journal} {Journal of Molecular Spectroscopy}\ }\textbf {\bibinfo
  {volume} {330}},\ \bibinfo {pages} {158} (\bibinfo {year}
  {2016}{\natexlab{b}})}\BibitemShut {NoStop}%
\bibitem [{\citenamefont {Fetter}\ and\ \citenamefont
  {Walecka}(2003)}]{Fetter2003}%
  \BibitemOpen
  \bibfield  {author} {\bibinfo {author} {\bibfnamefont {A.~L.}\ \bibnamefont
  {Fetter}}\ and\ \bibinfo {author} {\bibfnamefont {J.~D.}\ \bibnamefont
  {Walecka}},\ }\href
  {https://www.ebook.de/de/product/2192197/alexander_l_fetter_quantum_theory_of_many_particle_sys.html}
  {\emph {\bibinfo {title} {Quantum Theory of Many-Particle Systems}}}\
  (\bibinfo  {publisher} {Dover Publications Inc.},\ \bibinfo {year}
  {2003})\BibitemShut {NoStop}%
\bibitem [{\citenamefont {Jordan}\ and\ \citenamefont
  {Wigner}(1928)}]{jor28:631}%
  \BibitemOpen
  \bibfield  {author} {\bibinfo {author} {\bibfnamefont {P.}~\bibnamefont
  {Jordan}}\ and\ \bibinfo {author} {\bibfnamefont {E.}~\bibnamefont
  {Wigner}},\ }\href {\doibase 10.1007/bf01331938} {\bibfield  {journal}
  {\bibinfo  {journal} {Zeitschrift f{\"u}r Physik}\ }\textbf {\bibinfo
  {volume} {47}},\ \bibinfo {pages} {631} (\bibinfo {year} {1928})}\BibitemShut
  {NoStop}%
\bibitem [{\citenamefont {Hitchcock}(1927)}]{Hit27:164}%
  \BibitemOpen
  \bibfield  {author} {\bibinfo {author} {\bibfnamefont {F.~L.}\ \bibnamefont
  {Hitchcock}},\ }\href {\doibase 10.1002/sapm192761164} {\bibfield  {journal}
  {\bibinfo  {journal} {Journal of Mathematics and Physics}\ }\textbf {\bibinfo
  {volume} {6}},\ \bibinfo {pages} {164} (\bibinfo {year} {1927})}\BibitemShut
  {NoStop}%
\bibitem [{\citenamefont {Harshman}(1970)}]{har70:1}%
  \BibitemOpen
  \bibfield  {author} {\bibinfo {author} {\bibfnamefont {R.~A.}\ \bibnamefont
  {Harshman}},\ }\href {https://cir.nii.ac.jp/crid/1573387450564500608}
  {\bibfield  {journal} {\bibinfo  {journal} {UCLA Working Papers in
  Phonetics}\ }\textbf {\bibinfo {volume} {16}},\ \bibinfo {pages} {1}
  (\bibinfo {year} {1970})}\BibitemShut {NoStop}%
\bibitem [{\citenamefont {Carroll}\ and\ \citenamefont
  {Chang}(1970)}]{Car70:238}%
  \BibitemOpen
  \bibfield  {author} {\bibinfo {author} {\bibfnamefont {J.~D.}\ \bibnamefont
  {Carroll}}\ and\ \bibinfo {author} {\bibfnamefont {J.-J.}\ \bibnamefont
  {Chang}},\ }\href {\doibase 10.1007/bf02310791} {\bibfield  {journal}
  {\bibinfo  {journal} {Psychometrika}\ }\textbf {\bibinfo {volume} {35}},\
  \bibinfo {pages} {283} (\bibinfo {year} {1970})}\BibitemShut {NoStop}%
\bibitem [{\citenamefont {Kiers}(2000)}]{Kie00:105}%
  \BibitemOpen
  \bibfield  {author} {\bibinfo {author} {\bibfnamefont {H.~A.~L.}\
  \bibnamefont {Kiers}},\ }\href {\doibase
  10.1002/1099-128x(200005/06)14:3<105::aid-cem582>3.0.co;2-i} {\bibfield
  {journal} {\bibinfo  {journal} {Journal of Chemometrics}\ }\textbf {\bibinfo
  {volume} {14}},\ \bibinfo {pages} {105} (\bibinfo {year} {2000})}\BibitemShut
  {NoStop}%
\bibitem [{\citenamefont {Slater}(1929)}]{Sla29:1293}%
  \BibitemOpen
  \bibfield  {author} {\bibinfo {author} {\bibfnamefont {J.~C.}\ \bibnamefont
  {Slater}},\ }\href {\doibase 10.1103/physrev.34.1293} {\bibfield  {journal}
  {\bibinfo  {journal} {Physical Review}\ }\textbf {\bibinfo {volume} {34}},\
  \bibinfo {pages} {1293} (\bibinfo {year} {1929})}\BibitemShut {NoStop}%
\bibitem [{\citenamefont {Condon}(1930)}]{Con30:1121}%
  \BibitemOpen
  \bibfield  {author} {\bibinfo {author} {\bibfnamefont {E.~U.}\ \bibnamefont
  {Condon}},\ }\href {\doibase 10.1103/physrev.36.1121} {\bibfield  {journal}
  {\bibinfo  {journal} {Physical Review}\ }\textbf {\bibinfo {volume} {36}},\
  \bibinfo {pages} {1121} (\bibinfo {year} {1930})}\BibitemShut {NoStop}%
\bibitem [{\citenamefont {Kossaifi}\ \emph {et~al.}(2019)\citenamefont
  {Kossaifi}, \citenamefont {Panagakis}, \citenamefont {Anandkumar},\ and\
  \citenamefont {Pantic}}]{tensorly}%
  \BibitemOpen
  \bibfield  {author} {\bibinfo {author} {\bibfnamefont {J.}~\bibnamefont
  {Kossaifi}}, \bibinfo {author} {\bibfnamefont {Y.}~\bibnamefont {Panagakis}},
  \bibinfo {author} {\bibfnamefont {A.}~\bibnamefont {Anandkumar}}, \ and\
  \bibinfo {author} {\bibfnamefont {M.}~\bibnamefont {Pantic}},\ }\href
  {http://jmlr.org/papers/v20/18-277.html} {\bibfield  {journal} {\bibinfo
  {journal} {Journal of Machine Learning Research}\ }\textbf {\bibinfo {volume}
  {20}},\ \bibinfo {pages} {1} (\bibinfo {year} {2019})}\BibitemShut {NoStop}%
\bibitem [{\citenamefont {Kolda}\ and\ \citenamefont
  {Bader}(2009)}]{kol09:455}%
  \BibitemOpen
  \bibfield  {author} {\bibinfo {author} {\bibfnamefont {T.~G.}\ \bibnamefont
  {Kolda}}\ and\ \bibinfo {author} {\bibfnamefont {B.~W.}\ \bibnamefont
  {Bader}},\ }\href {\doibase 10.1137/07070111X} {\bibfield  {journal}
  {\bibinfo  {journal} {SIAM Rev}\ }\textbf {\bibinfo {volume} {51}},\ \bibinfo
  {pages} {455} (\bibinfo {year} {2009})}\BibitemShut {NoStop}%
\bibitem [{\citenamefont {Worth}\ \emph {et~al.}()\citenamefont {Worth},
  \citenamefont {Beck}, \citenamefont {J{\"a}ckle}, \citenamefont {Vendrell},\
  and\ \citenamefont {Meyer}}]{mctdh:MLpackage}%
  \BibitemOpen
  \bibfield  {author} {\bibinfo {author} {\bibfnamefont {G.~A.}\ \bibnamefont
  {Worth}}, \bibinfo {author} {\bibfnamefont {M.~H.}\ \bibnamefont {Beck}},
  \bibinfo {author} {\bibfnamefont {A.}~\bibnamefont {J{\"a}ckle}}, \bibinfo
  {author} {\bibfnamefont {O.}~\bibnamefont {Vendrell}}, \ and\ \bibinfo
  {author} {\bibfnamefont {H.-D.}\ \bibnamefont {Meyer}},\ }\href@noop {}
  {}\bibinfo {howpublished} {The {MCTDH} {P}ackage, {V}ersion 8.2, (2000).
  H.-D. Meyer, {V}ersion 8.3 (2002), {V}ersion 8.4 (2007). O. Vendrell and
  H.-D. Meyer {V}ersion 8.5 (2013). {V}ersion 8.5 contains the ML-MCTDH
  algorithm. Current versions: 8.4.18 and 8.5.11 (2019). Used version: 
  exchange with "Used version" {S}ee http://mctdh.uni-hd.de/}\BibitemShut
  {NoStop}%
\end{thebibliography}
%
\end{document}